\documentclass[twocolumn,tighten,times,twocolappendix,numberedappendix]{aastex631}

\usepackage{amsmath}
\usepackage{float}
\usepackage{color}
\usepackage{url}
\usepackage{etoolbox}
\usepackage{graphicx}

\usepackage[flushleft]{threeparttable}

\hypersetup{breaklinks=true,colorlinks=true,linkcolor=blue,citecolor=blue,filecolor=blue,urlcolor=blue}
\makeatletter
\patchcmd{\NAT@citex}
  {\@citea\NAT@hyper@{%
     \NAT@nmfmt{\NAT@nm}%
     \hyper@natlinkbreak{\NAT@aysep\NAT@spacechar}{\@citeb\@extra@b@citeb}%
     \NAT@date}}
  {\@citea\NAT@nmfmt{\NAT@nm}%
   \NAT@aysep\NAT@spacechar\NAT@hyper@{\NAT@date}}{}{}

\patchcmd{\NAT@citex}
  {\@citea\NAT@hyper@{%
     \NAT@nmfmt{\NAT@nm}%
     \hyper@natlinkbreak{\NAT@spacechar\NAT@@open\if*#1*\else#1\NAT@spacechar\fi}%
       {\@citeb\@extra@b@citeb}%
     \NAT@date}}
  {\@citea\NAT@nmfmt{\NAT@nm}%
   \NAT@spacechar\NAT@@open\if*#1*\else#1\NAT@spacechar\fi\NAT@hyper@{\NAT@date}}
  {}{}

\usepackage{array}
\newcolumntype{C}[1]{>{\centering\let\newline\\\arraybackslash\hspace{0pt}}m{#1}}

\usepackage{xspace}

\newcommand{\orcidicon}{\includegraphics[width=0.26cm]{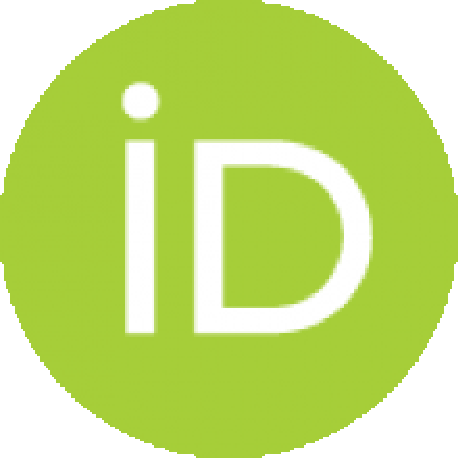}}

\newcommand{\orcidauthor}[1]{\href{https://orcid.org/#1}{\orcidicon}}

\def\apj{ApJ}
\def\apjl{ApJ}
\def\apjs{ApJS}
\def\ao{Appl.~Opt.}

\def\aap{A\&A}
\def\aapr{A\&A~Rev.}
\def\aaps{A\&AS}

\def\mnras{MNRAS}

\def\pasp{PASP}
\def\pasj{PASJ}

\def\nat{Nature}

\def\procspie{Proc.~SPIE}

\def\actaa{AcA}

\newcommand{\ionic}[2]{#1$\,${\scshape{#2}}\xspace}

\def\arcdeg{\hbox{$^\circ$}}

\makeatletter
\patchcmd{\frontmatter@RRAP@format}{(}{}{}{}
\patchcmd{\frontmatter@RRAP@format}{)}{}{}{}
\renewcommand\Dated@name{}
\makeatother

\renewcommand{\vec}[1]{\mathbf{#1}}

\shorttitle{X-ray Variability in RT\,Cru: PCA}
\shortauthors{Danehkar et al.}
\graphicspath{{./}{figures/}}

\begin{document}

\title{X-ray Variability in the Symbiotic Binary RT\,Cru: Principal Component Analysis}

\correspondingauthor{A.~Danehkar}

\author[0000-0003-4552-5997]{A.~Danehkar}
\affiliation{Eureka Scientific, 2452 Delmer Street, Suite 100, Oakland, CA 94602-3017, USA; \href{mailto:danehkar@eurekasci.com}{danehkar@eurekasci.com}
}

\author[0000-0002-0210-2276]{J.~J.~Drake}
\affiliation{Lockheed Martin, 3251 Hanover St, Palo Alto, CA 94304, USA}

\author[0000-0002-2647-4373]{G.~J.~M.~Luna}
\affiliation{CONICET-Universidad Nacional de Hurlingham, Av. Gdor. Vergara 2222, Villa Tesei, Buenos Aires, Argentina
}



\date[]{\textit{\footnotesize Received 2023 December 31; revised 2024 June 22; accepted 2024 June 23; published 2024 August 28}}

\begin{abstract}
Hard X-ray-emitting ($\delta$-type) symbiotic binaries, which exhibit a strong hard X-ray excess, have posed a challenge to our understanding of accretion physics in degenerate dwarfs. RT\,Cru, which is a member of the $\delta$-type symbiotics, shows stochastic X-ray variability. Timing analyses of X-ray observations from \textit{XMM-Newton} and \textit{NuSTAR}, which we consider here, indicate hourly fluctuations, in addition to a spectral transition from 2007 to a harder state in 2012 seen with \textit{Suzaku} observations. To trace the nature of X-ray variability, we analyze the multi-mission X-ray data using principal component analysis (PCA), which determines the spectral components that contribute most to the flickering behavior and the hardness transition. The \textit{Chandra} HRC-S/LETG and \textit{XMM-Newton} EPIC-pn data provide the primary PCA components, which may contain some variable emission features, especially in the soft excess. Additionally, the absorbing column (first order with 50\%), along with the source continuum (20\%), and a third component (9\%) -- which likely accounts for thermal emission in the soft band -- are the three principal components found in the \textit{Suzaku} XIS1 observations. The PCA components of the \textit{NuSTAR} data also correspond to the continuum and possibly emission features. Our findings suggest that the spectral hardness transition between the two \textit{Suzaku} observations is mainly due to changes in the absorbing material and X-ray continuum, while some changes in the thermal plasma emission may result in flickering-type variations.
\end{abstract}

\keywords{\href{https://astrothesaurus.org/uat/1674}{Symbiotic binary stars (1674)};
\href{https://astrothesaurus.org/uat/1578}{Stellar accretion (1578)};
\href{https://astrothesaurus.org/uat/1822}{X-ray sources (1822)};
\href{https://astrothesaurus.org/uat/1916}{Time series analysis (1916)};
\href{https://astrothesaurus.org/uat/1944}{Principal component analysis (1944)}
\vspace{4pt}
}


\section{Introduction}
\label{rtcru:introduction}



Symbiotic systems refer to binary stars that are characterized by the presence of a hot degenerate core accreting matter from a cool red giant star \citep{Paczynski1980,Kenyon1984,Belczynski2000}. They exhibit soft or supersoft thermal X-ray emission \citep[][]{Muerset1997,Luna2013} dominated by blackbody-like or bremsstrahlung radiation \citep{Imamura1983}. However, a small group of them have been observed to emit an extreme hard X-ray excess above 2.4\,keV \citep{Tueller2005,Bird2007,Kennea2009,Eze2014}. These are the  \textit{hard X-ray-emitting} symbiotics or $\delta$-type sources according to the classification scheme by \citet{Luna2013}. They could be progenitors of type Ia supernovae owing to the possible presence of massive white dwarfs \citep{Kennea2009}. Among this group, notable systems can be mentioned: RT\,Cru \citep{Luna2007,Ducci2016,Luna2018,Danehkar2021}, CH\,Cyg \citep{Wheatley2006,Mukai2007,Toala2023}, T\,CrB \citep{Luna2008,Zhekov2019}, SS73\,17 \citep[CD--57\,3057;][]{Smith2008,Eze2010}, and MWC\,560 \citep{Stute2009,Lucy2020}. The discovery of this particular group presents a challenge to our knowledge about accretion physics in white dwarfs due to the strong hard X-ray emission \citep{Luna2007,Kennea2009}. 
Recently, \citet{Toala2024} proposed a disk-like model to explain the X-ray properties of symbiotics, where $\delta$-type sources have an accretion disk near the edge. In addition, radiative transfer simulations of X-ray photons by \citet{Toala2024} implied that the $\delta$-type group is likely related to low-accreting degenerate cores with high-temperature plasma ($> 1$\,keV) within the boundary layer between the inner edge of the accretion disk and the white dwarf surface.

\begin{table*}
\begin{center}
\caption[]{Observation log of RT\,Cru.
\label{rtcru:obs:log}}
\footnotesize
\begin{tabular}{llllccrrr}
\hline\hline\noalign{\smallskip}
Observatory & Instrument     & Config. & Obs. ID      & \multicolumn{1}{c}{Obs. Start (UTC)}     & \multicolumn{1}{c}{Obs. End (UTC)}        & Exp. (ks)\,$^{\rm \bf a}$  & Count\,$^{\rm \bf a}$      & Cnt.\,Rate\,$^{\rm \bf a}$ \\
\noalign{\smallskip}
\tableline
\noalign{\smallskip}
\textit{Suzaku}      & XIS1           & Pointing           & 402040010    & 2007 Jul 02, 12:38   & 2007 Jul 03, 05:50    &     50.88 &       35043 &      0.689 \\
\noalign{\smallskip}
\textit{Suzaku}      & HXD-PIN            & Pointing           & 402040010    & 2007 Jul 02, 12:38   & 2007 Jul 03, 05:50    &     40.17 &       30126 &      0.750 \\
\noalign{\smallskip}
\textit{Suzaku}      & XIS1           & Pointing           & 906007010    & 2012 Feb 06, 18:17   & 2012 Feb 07, 20:00    &     39.43 &       25291 &      0.641 \\
\noalign{\smallskip}
\textit{Suzaku}      & HXD-PIN            & Pointing           & 906007010    & 2012 Feb 06, 18:17   & 2012 Feb 07, 20:00    &     32.55 &       20196 &      0.620 \\
\noalign{\smallskip}
\textit{Chandra}     & HRC-S/LETG     &                    & 16688        & 2015 Nov 23, 02:01   & 2015 Nov 23, 09:38    &     25.15 &        3859 &      0.153 \\
\noalign{\smallskip}
\textit{Chandra}     & HRC-S/LETG     &                    & 18710        & 2015 Nov 23, 22:42   & 2015 Nov 24, 14:13    &     53.73 &        8313 &      0.155 \\
\noalign{\smallskip}
\textit{NuSTAR}      & FPMA+B         &                    & 30201023002  & 2016 Nov 20, 00:41   & 2016 Nov 21, 02:46    &     58.21 &      117066 &      2.012 \\
\noalign{\smallskip}
\textit{XMM}         & EPIC-pn        & Imaging    & 0831790801   & 2019 Mar 03, 05:39   & 2019 Mar 03, 21:05    &     47.91\,[40.75] &        2415\,[1997] &      0.049 \\
\noalign{\smallskip}\hline
\end{tabular}
\end{center}
\begin{tablenotes}
\footnotesize
\item[1]\textbf{Note.} $^{\rm \bf a}$ Source counts and count-rates over 0.4--10\,keV for \textit{Suzaku} (XIS1), \textit{XMM-Newton} (EPIC-pn) and \textit{Chandra} 
(HRC-S/LETG: LEG $m = \pm 1$), 10--70\,keV for \textit{Suzaku} (HXD-PIN), and 3--79\,keV for \textit{NuSTAR} (FPMA+B).
The data in the square brackets correspond to the \textit{XMM-Newton} events without flaring background.
\end{tablenotes}
\end{table*}


Some hard X-ray-emitting symbiotics seem to produce distinctive \textit{soft} and \textit{hard} thermal plasma emissions: CH\,Cyg with plasma temperatures of 0.2, 0.7, and 7.3\,keV \citep[][]{Ezuka1998}, SS73\,17 with temperatures of $1.12$ and $9.9$\,keV \citep[][]{Eze2010}, and MWC\,560 showing thermal emissions with temperatures of $0.18$ and $11.26$\,keV \citep[][]{Stute2009}. Symbiotic stars with distinctive soft and hard X-ray thermal components are referred to as the $\beta/\delta$-type group \citep{Luna2013}. Previous studies have also identified the presence of jets in some of them, namely CH\,Cyg \citep{Galloway2004,Karovska2007,Karovska2010} and MWC\,560 \citep{Tomov1992,Schmid2001,Lucy2018}. The soft thermal emission found in these systems might have a potential link to the shock region created by either a jet or the interaction of a wind colliding with the surrounding material \citep{Stute2009}. According to \citet{Toala2024}, the soft X-ray emission of a two-temperature plasma model can be obscured by the disk, so extended emission from jets, colliding winds, and/or hot bubbles is likely responsible for the soft component in $\beta/\delta$ sources. The soft X-ray component could originate from colliding stellar winds (CSWs), though a different mechanism such as accretion may also be responsible for the X-ray emission seen in some epochs in the $\beta$-class symbiotic star AG Peg \citep{Zhekov2016}.


Although the previous X-ray data of RT\,Cru revealed only highly absorbed, hard thermal plasma emission of 8.6 keV \citep{Luna2007}, a recent Bayesian statistical analysis of the latest \textit{Chandra} observations also suggested the possible existence of heavily obscured, soft plasma emission with a temperature of $\sim 1.3$ keV in addition to a hard thermal component with a temperature of 9.6 keV \citep{Danehkar2021}. However, a recent statistical method based on differences between the backgrounds from the smooth and likelihood-ratio tests could not robustly put constraints on low-energy emission lines of the soft thermal emission, but yielded an upper confidence limit of 1\,keV on the soft plasma temperature \citep{Zhang2023}. 
A dramatic decline in optical emission lines and hard X-ray emission in RT\,Cru observed to have occurred in 2019 was attributed to a strong decline in accretion activity \citep{Pujol2023}. Nevertheless, high X-ray variability of RT\,Cru indicates that some dense materials along the line of sight could potentially block a large portion of the thermal emission \citep{Danehkar2021}. Accordingly, the soft thermal emission component ($\sim 1$ keV) might be heavily obscured by such material, making it difficult to detect with significant statistics.

The X-ray variability of RT\,Cru recorded with the \textit{Chandra} telescope has been recently investigated using hardness ratio analysis and spectral modeling \citep{Danehkar2021}. To further evaluate the nature of the X-ray variability in this $\delta$-type symbiotic star and the soft thermal plasma emission suggested by \citet{Danehkar2021}, we conduct further timing analyses on the archival data taken with \textit{XMM-Newton}, \textit{Suzaku}, and \textit{NuSTAR}, in addition to comprehensive eigenvector-based multivariate analyses of the historical X-ray data of the \textit{Chandra} and other telescopes.
Section~\ref{rtcru:observation} describes our reduction of time-sliced data required for implementing principal component analysis (PCA). In Section~\ref{rtcru:timing}, we investigate the X-ray light curves and hardness ratios. Section~\ref{rtcru:pca} presents our principal components determined with PCA, as well as simulated PCA spectra, and is followed by discussions in Section~\ref{rtcru:discussions} and a conclusion in Section~\ref{rtcru:conclusion}.

\vfill\break

\section{Data Reduction for PCA}
\label{rtcru:observation}

RT\,Cru was observed using the X-ray Imaging Spectrometer \citep[XIS;][]{Koyama2007} and the Hard X-ray Detector \citep[HXD;][]{Takahashi2007} on board \textit{Suzaku} \citep[][]{Mitsuda2007} in 2007 and 2012, the EPIC-pn instrument \citep[][]{Strueder2001} aboard the \textit{XMM-Newton} telescope \citep[][]{Jansen2001} in 2019,  
the `A' and `B' focal plane modules (FPM) on the Nuclear Spectroscopic Telescope Array (\textit{NuSTAR}) satellite \citep[][]{Harrison2013} in 2016, and the Low Energy Transmission Grating \citep[LETG;][]{Brinkman2000} on the High Resolution Camera Spectrometer \citep[HRC-S;][]{Murray2000} of \textit{Chandra} X-ray Observatory \citep[CXO;][]{Weisskopf2000,Weisskopf2002} in 2015. The observations are summarized in Table~\ref{rtcru:obs:log}, which includes the instrument and its configuration, observation identification  number, start and end times (UTC), exposure time (ks), and total counts and count rate (count\,s$^{-1}$) of each observation. To implement eigenvector-based multivariate statistical analysis such as PCA of a variable X-ray source, it is necessary to split each dataset into a time series of spectral data at fixed intervals of 10\,ks.

We downloaded the \textit{Chandra} HRC-S/LETG event data from the Chandra data archive 
and reprocessed them using the \textsf{chandra\_repro} tool from the \textsc{ciao} package \citep[version 4.15;][]{Fruscione2006} together with the Chandra CALDB data (version 4.10.2).\footnote{The \textit{Chandra} dataset is contained in~\dataset[doi:10.25574/cdc.201]{https://doi.org/10.25574/cdc.201}.}
The time-segmented event files were produced by applying the \textsc{ciao} operations \textsf{dmcopy} and \textsf{dmappend} on the 2nd level events. The low-energy grating (LEG) spectra, together with their respective redistribution and response data, were generated by applying the \textsc{ciao} programs \textsf{tgextract} and \textsf{mktgresp} to the time-sliced event files. The \textsf{dmtype2split} tool, part of the \textsc{ciao} software, was utilized to segregate distinct positive and negative orders, 
while the application \textsf{tg\_bkg} was used to create the background spectra.

\begin{figure*}
\begin{center}
\includegraphics[height=9.70cm, trim = 0 0 0 0, clip, angle=0]{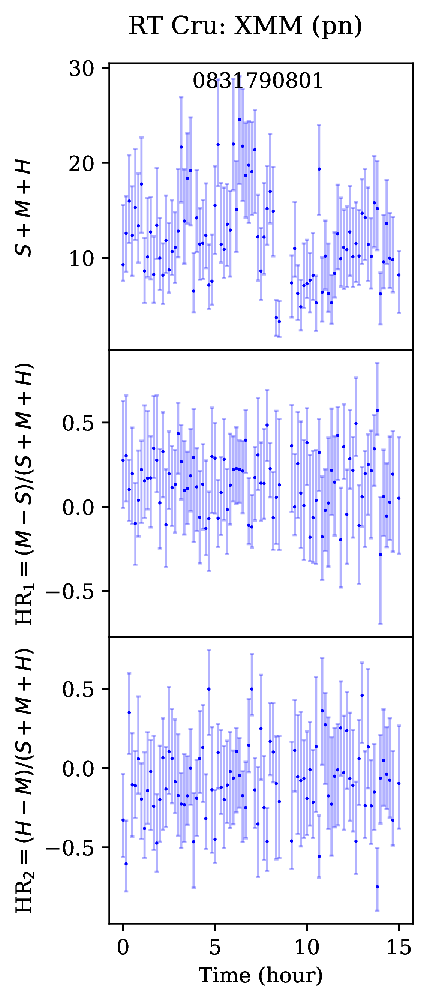}%
\includegraphics[height=9.70cm, trim = 0 0 0 0, clip, angle=0]{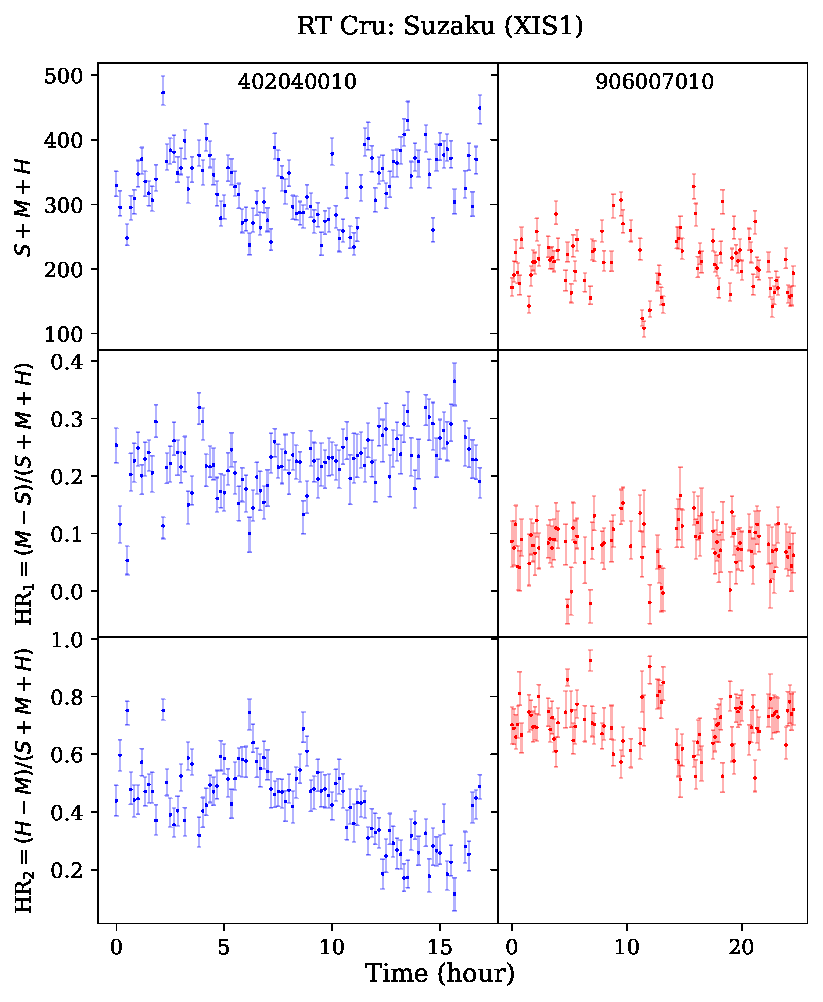}\\
\includegraphics[height=12.86cm, trim = 0 0 0 0, clip, angle=0]{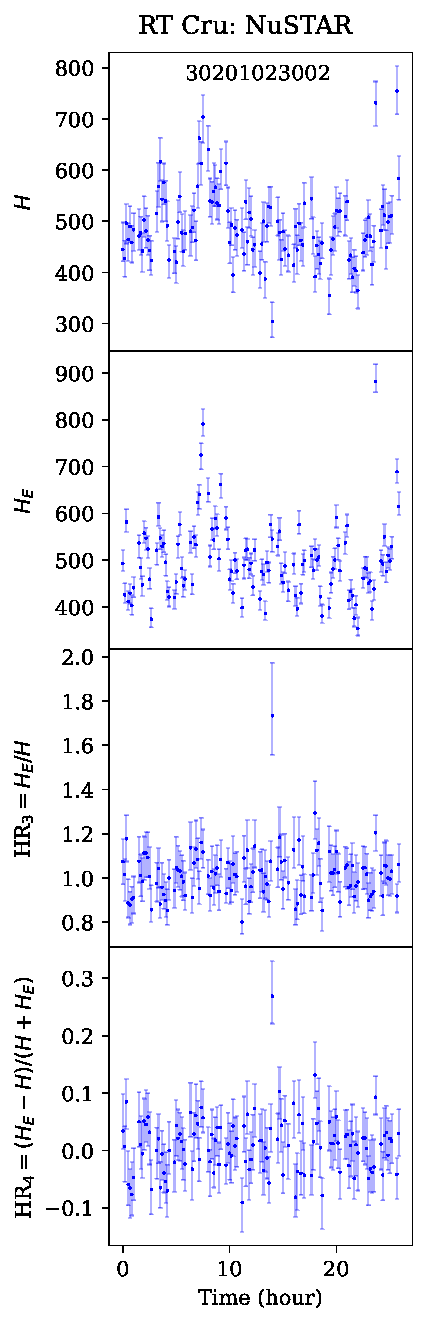}%
\includegraphics[height=12.86cm, trim = 0 0 0 0, clip, angle=0]{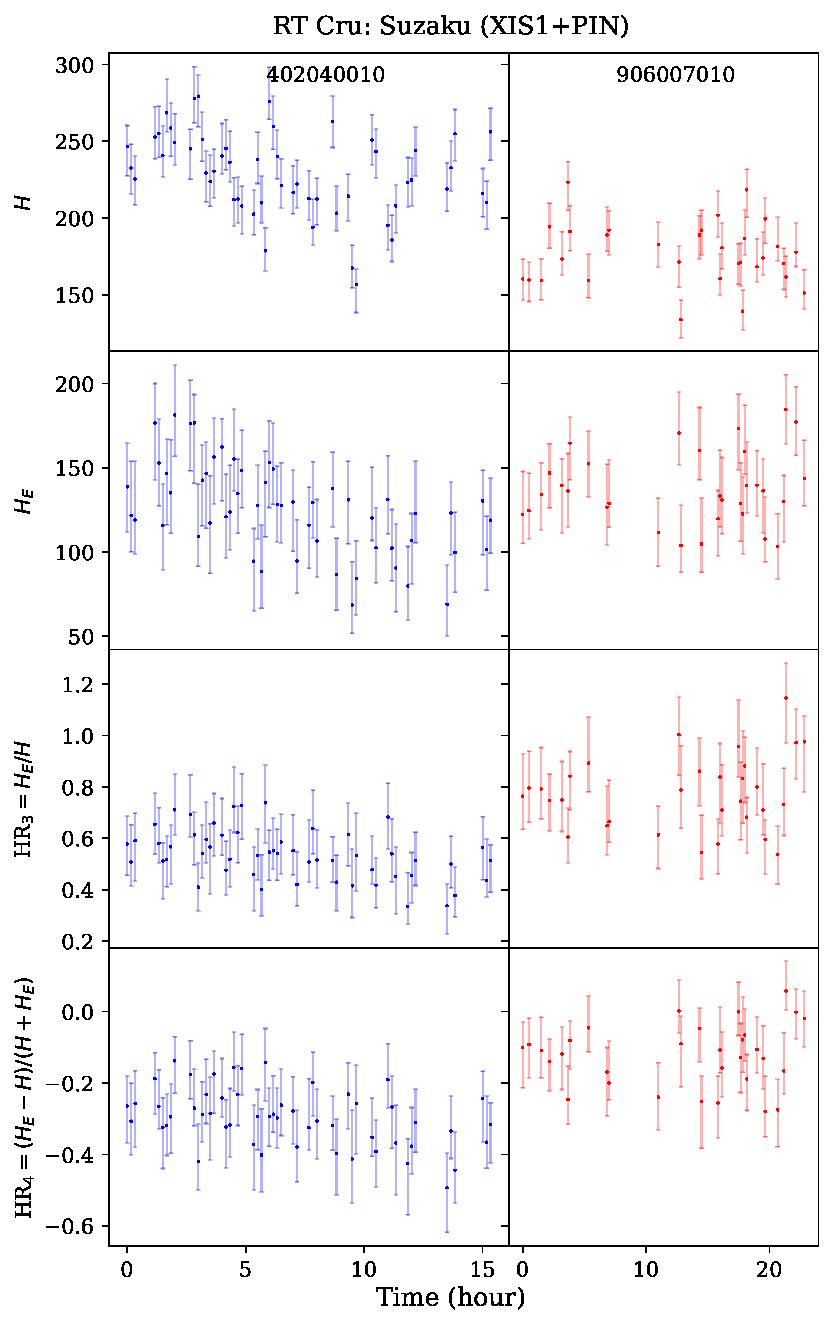}%
\end{center}
\caption{The background-subtracted light curves of RT\,Cru in the energy bands $S+M+H$, $H$ and $H_{E}$ (in counts) binned at 600 sec, along with the corresponding hardness ratios ${\rm HR}_{1} = (M-S)/(S+M+H)$, ${\rm HR}_{2} = (H-M)/(S+M+H)$, ${\rm HR}_{\rm 3} = H_{E}/H$ and ${\rm HR}_{\rm 4} = (H_{E}-H)/(H+H_{E})$ computed with the BEHR using the source and background time series of the \textit{XMM-Newton} (EPIC-pn), \textit{Suzaku} (XIS1), \textit{NuSTAR}, and \textit{Suzaku} (XIS1+PIN) observations. The second \textit{Suzaku} light curves were amplified according to the instrument sensitivities with respect to the first one.
\label{rtcru:fig:lc}
}
\end{figure*}

The \textit{XMM-Newton} EPIC-pn data were obtained from XMM Science Archive and reduced with the science analysis software \citep[\textsc{sas} v\,20.0.0;][]{Gabriel2004} and the calibration files (XMM-CCF-REL-391). 
The use of the \textsc{sas} tool \textsf{epproc} led to the generation of processed event files. These events were subsequently employed to produce new event files, which were stacked at intervals of 10\,ks with the \textsc{sas} program \textsf{evselect}. We removed the events of flaring particle background, which were identified using count rates of $> 0.4$ c\,s$^{-1}$ in the single-pixel (PATTERN\,$=$\,0) light curves binned at 100\,sec within the 10--12\,keV energy range. Furthermore, the time-filtered event files exclusively included single and double patterned events (PATTERN\,$\leqslant$\,4) within the relevant pulse-invariant (PI) channel range (200\,$<$\,PI\,$<$\,15000), disregarding defective pixels (FLAG\,$=$\,0). The procedure \textsf{especget} was utilized to generate a set of source spectra by applying it to the time-sliced event data. The spectra of the source were taken from a circular region with a radius of $36''$ centered on the brightest peak of the source. The background spectra were created using a circle of the same size on the same chip, but without any sources included.

The \textit{Suzaku} data acquired using the XIS and HXD-PIN were retrieved and processed using the \textsc{ftool} program \textsf{aepipeline} from the HEAsoft software (v\,6.31.1) 
and the calibration data (XIS: 2018-10-10 and HXD: 2011-09-13). 
The XIS provided energy coverage ranging from 0.2 to 12 keV, while the HXD supplied a band pass of 10-70 keV using PIN diodes. Various Good Time Interval (GTI) files at our chosen time interval were created using the \textsc{ftool} application \textsf{maketime}. The GTI files were employed in the multipurpose tool \textsc{xselect} to generate a set of time-segmented source spectra.  
The source spectrum was acquired by extracting data from a circular region with a radius of $216''$ centered on the emission peak of the source, whereas the background was chosen from a nearby circular region of the same radius excluding any sources. We should specify that XIS1 was the only back-illuminated (BI) detector aboard \textit{Suzaku}, whereas the other XIS devices 
were front-illuminated (FI). 
Typically, BI detectors are expected to offer superior quantum efficiency to FI detectors for an identical depletion depth \citep{Lesser1998}. 
The program \textsf{hxdpinxbpi} was employed to generate the time-sliced spectra of HXD-PIN data using the GTI files and the "tuned" non-X-ray PIN background released by the \textit{Suzaku} team.\footnote{\href{https://heasarc.gsfc.nasa.gov/FTP/suzaku/data/background/}{https://heasarc.gsfc.nasa.gov/FTP/suzaku/data/background/}}

The \textit{NuSTAR} data taken with the FPM `A' and `B' were downloaded from the HEASARC archive 
and reduced using the tool \textsf{nupipeline} from NuSTARDAS (v\,2.1.2) and the relevant calibration files (8 Feb 2023). The \textsc{ftool} program \textsf{maketime} was used to create the GTI tables at intervals of 10\,ks, which were then applied to the calibrated event files via the application \textsf{nuproducts} resulting in the time-stacked spectra. The source was extracted from a circular aperture with a radius of $75''$, whereas the background was from a location on the same chip that was devoid of any sources.

\section{Time Series Analysis }
\label{rtcru:timing}

To conduct timing analysis, we produced light curves in a variety of energy bands for the \textit{Suzaku} (XIS1, PIN), \textit{XMM-Newton} (pn), and \textit{NuSTAR} data, which helped us identify the spectral transitions in the X-ray observations of RT\,Cru. We chose the soft ($S$: 0.4--1.1\,keV), medium ($M$: 1.1--2.6\,keV), hard ($H$: 2.6--10\,keV), and extreme hard ($H_{E}$: 10--50\,keV) bands, apart from the hard band of 3--10\,keV for \textit{NuSTAR} data. To enhance signal-to-noise ratios, we discretized the time series into 600-second binning intervals, which have enough temporal resolution to distinguish any spectral variations happening on hourly timescales. The \textit{XMM-Newon} light curves were generated in the desired time bins and energy ranges with the \textsc{sas} program \textsf{evselect}, while the time-binned light curves of the \textit{Suzaku} XIS1 data were generated using the typical filtering techniques in \textsc{xselect}. The \textit{NuSTAR} light curves were also created using the application \textsf{nuproducts}. In the program \textsc{xselect}, we also created the time-filtered HXD/PIN events, which were then used by the \textsc{ftool} task \textsf{hxdpinxblc} to build the light curves of the \textit{Suzaku} PIN data.
To correct for the decline in the instrument sensitivities with time, the \textit{Suzaku} light curves in 2012 were scaled up in relation to the 2007 observation. This scaling was decided based on the integration of the effective areas over the given energy range, $\int_{E=E_{\rm min}}^{E_{\rm max}} A_{\rm eff} (E) dE$, where $A_{\rm eff}(E)$ is the effective function from the auxiliary response file (ARF), and $E_{\rm min}$ and $E_{\rm max}$ are the lower and upper limits of the band, respectively.

The time series produced in different energy bands allow us to investigate spectral evolution over time. To distinguish different spectral states similar to what was done for the \textit{Chandra} data \citep{Danehkar2021}, the following hardness ratios are computed using the time-binned light curves from the four energy bands:
\begin{align}
{\rm HR}_{1} = \frac{M-S}{S+M+H},~~~~~ & {\rm HR}_{2} = \frac{H-M}{S+M+H},\label{eq_1}\\
{\rm HR}_{3} = \frac{H_{E}}{H},~~~~~ & {\rm HR}_{4} = \frac{H_{E}-H}{H+H_{E}}.\label{eq_2}
\end{align}
The first two equations are identical to those employed by \citet{Prestwich2003} for the classification of X-ray sources in the Local Group galaxies. The hardness ratio diagrams are created by plotting the hardness ratios against the entire bands, and assist in the detection of spectral transitions associated with accretion or obscuration caused by absorbing material. Hardness ratio analysis has been employed to characterize various astronomical objects, including extragalactic X-ray sources \citep[e.g.,][]{Hong2004,Plucinsky2008}, quasars \citep[][]{Danehkar2018,Boissay-Malaquin2019}, and X-ray binaries \citep{Sreehari2021}.

\begin{figure*}
\begin{center}
\includegraphics[height=11.2cm, trim = 0 0 0 0, clip, angle=0]{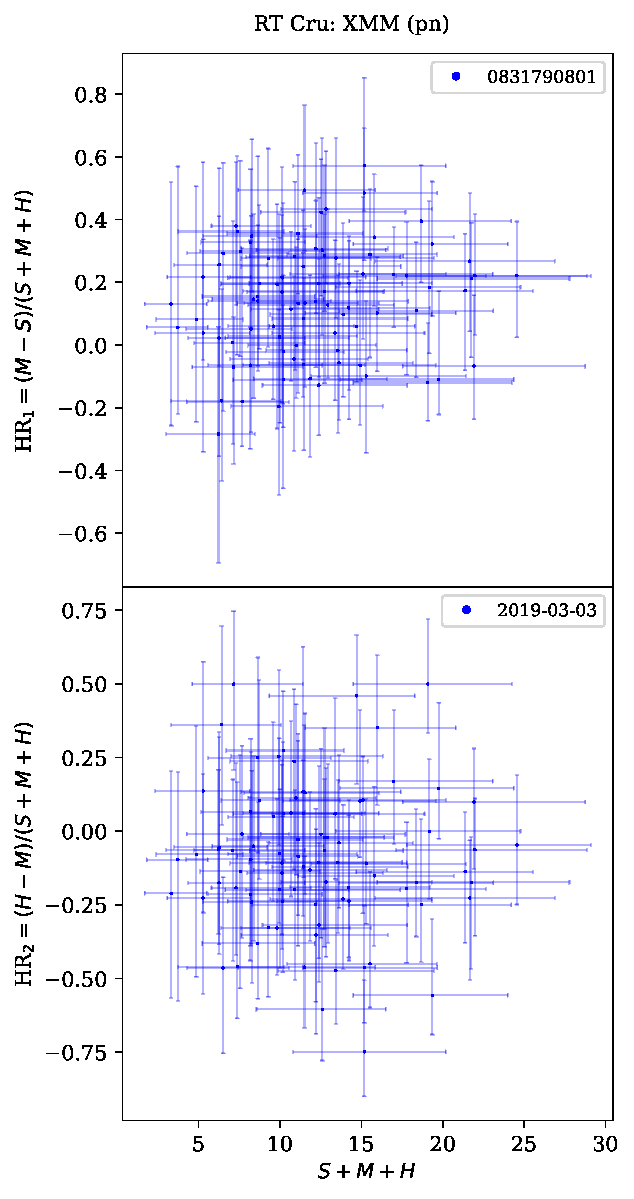}%
\includegraphics[height=11.2cm, trim = 0 0 0 0, clip, angle=0]{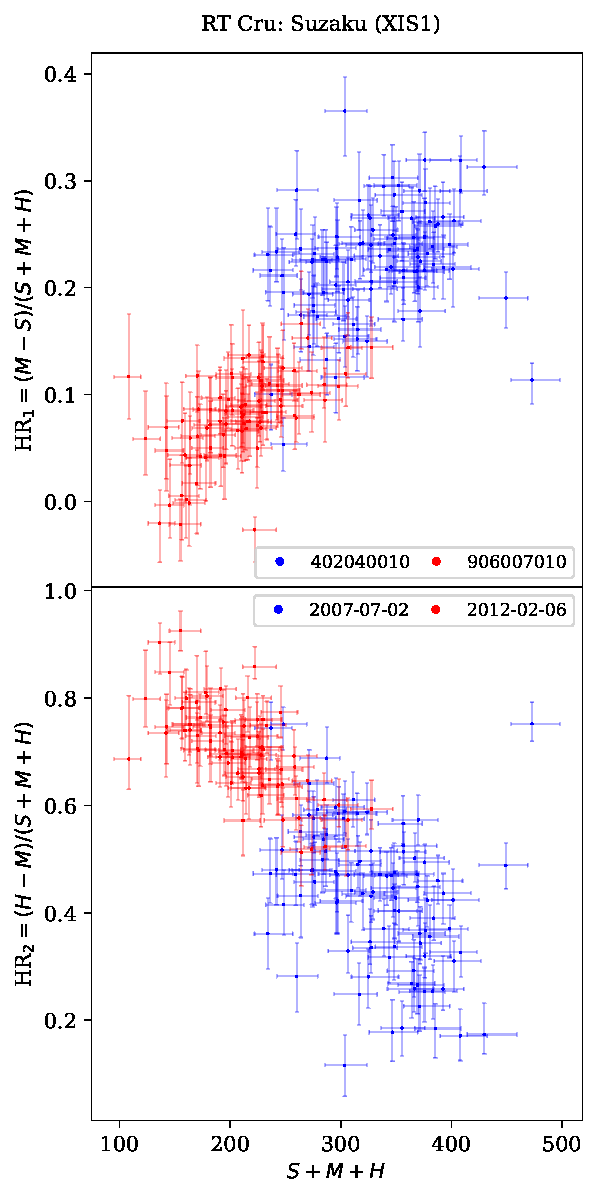}\\
\includegraphics[height=11.2cm, trim = 0 0 0 0, clip, angle=0]{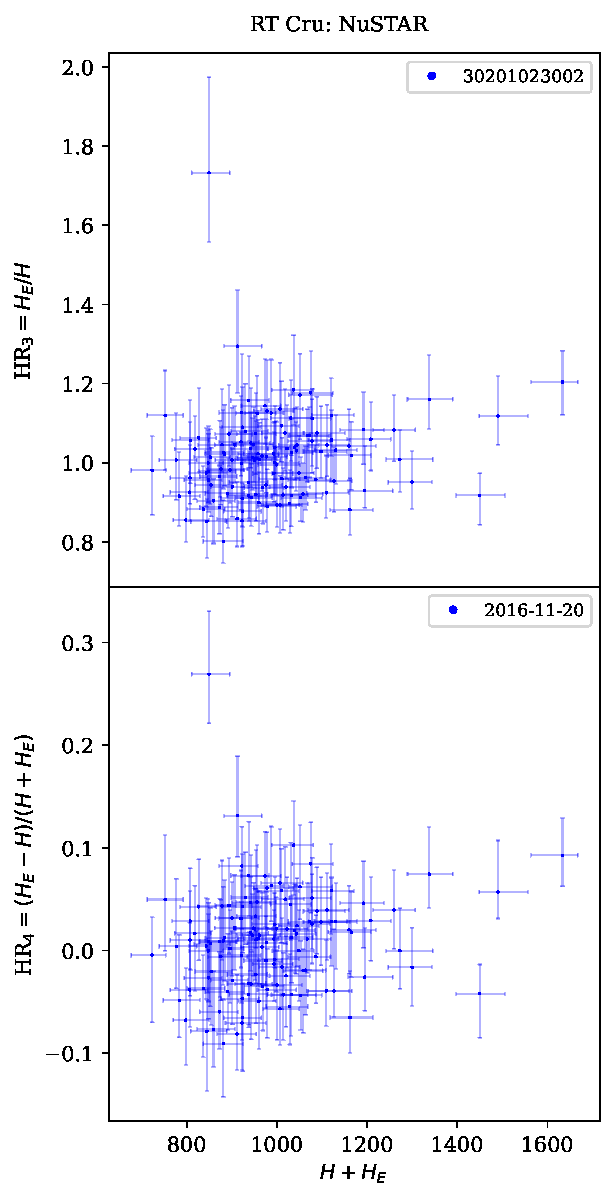}%
\includegraphics[height=11.2cm, trim = 0 0 0 0, clip, angle=0]{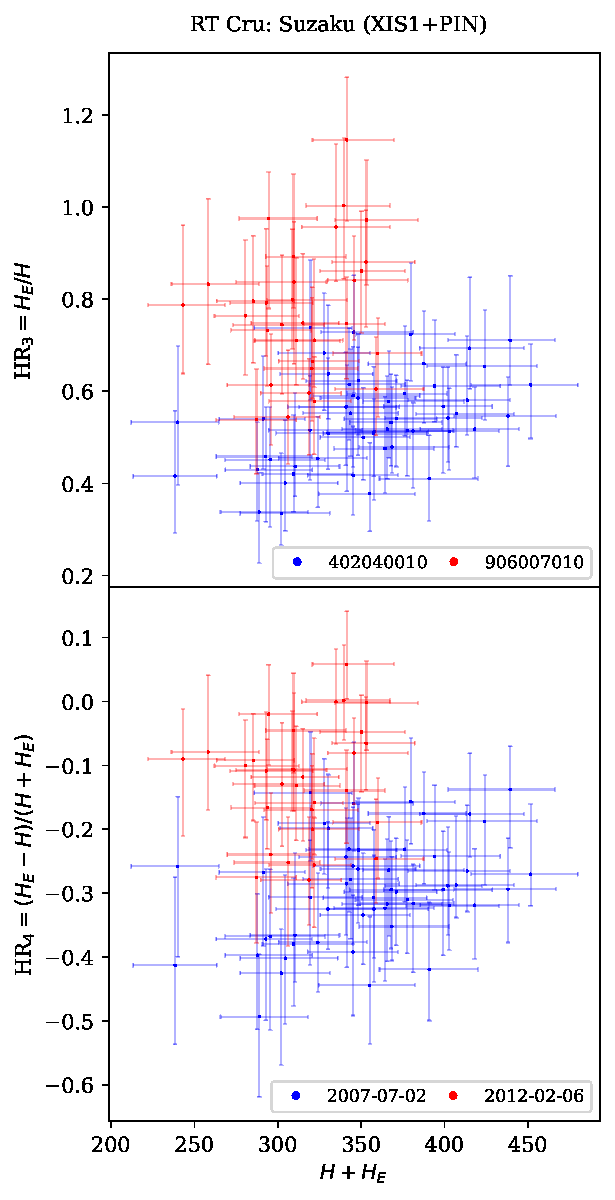}%
\end{center}
\caption{The hardness ratio diagrams of RT\,Cru: ${\rm HR}_{1} = (M-S)/(S+M+H)$ and ${\rm HR}_{2} = (H-M)/(S+M+H)$  plotted against $S+M+H$ (in counts), ${\rm HR}_{\rm 3} = H_{E}/H$ and ${\rm HR}_{\rm 4} = (H_{E}-H)/(H+H_{E})$  plotted against $H+H_{E}$ (in counts) computed with the BEHR using the 600\,s-binned source and background light curves of the \textit{XMM-Newton} (EPIC-pn), \textit{Suzaku} (XIS1), \textit{NuSTAR}, and \textit{Suzaku} (XIS1+PIN) observations. The second \textit{Suzaku} counts were magnified based on the instrument sensitivities relative to the first one.
\label{rtcru:fig:hdr}
}
\end{figure*}

In Figure~\ref{rtcru:fig:lc}, we present the background-subtracted light curves of RT\,Cru made with the \textit{XMM-Newton}, \textit{Suzaku}, and \textit{NuSTAR} observations binned at intervals of 600\,sec in the broad ($S+M+H$; 0.4--10\,keV), hard ($H$), and extreme hard ($H_{E}$) bands. The figure also shows the corresponding hardness ratios, namely ${\rm HR}_{1}$ and ${\rm HR}_{2}$ for \textit{XMM-Newton} (pn) and \textit{Suzaku} (XIS1), ${\rm HR}_{\rm 3}$ and ${\rm HR}_{\rm 4}$ for \textit{NuSTAR} and \textit{Suzaku} (XIS1+PIN). The uncertainties in the light curves and hardness ratios were determined using the Bayesian Estimator for Hardness Ratios \citep[BEHR;][]{Park2006}. Hour-scale variations are evident in all the time series. In particular, the source got harder in 2012 February compared to 2007 July according to the ${\rm HR}_{2}$ time series of the \textit{Suzaku} XIS1 data, whereas the source brightness ($S+M+H$) in 2012 is lower than that in 2007. Moreover, the ${\rm HR}_{\rm 3}$ and ${\rm HR}_{\rm 4}$ hardness ratios involving the extreme hard band ($H_{E}$) were stronger in 2012 February.

Figure~\ref{rtcru:fig:hdr} shows hardness ratio diagrams that illustrate the relationship between the hardness ratios and the broad bands ($S+M+H$ and $H+H_{E}$). The hardness diagrams of the \textit{XMM-Newton} observation do not depict a hardness transition similar to what was seen with the \textit{Chandra} data \citep{Danehkar2021}. However, the $\mathrm{HR}_{1}$ and $\mathrm{HR}_{2}$ diagrams of the two \textit{Suzaku} observations indicate that the source experienced a transition from the high/soft to low/hard spectral states between 2007 and 2012. Moreover, the \textit{Suzaku} diagrams also revealed that the source was stronger in the soft (0.4--1.1\,keV) and hard (2.6--10\,keV) bands -- lower $\mathrm{HR}_{1}$ and higher $\mathrm{HR}_{2}$ at the same time -- in 2012 February by comparison with those in 2007 July. Although we do not see any special pattern in the hardness ratio diagrams of \textit{NuSTAR}, the $\mathrm{HR}_{3}$ and $\mathrm{HR}_{4}$ diagrams of the \textit{Suzaku} light curves imply a higher average of the extreme hardness ($H_{E}$) in 2012 February.

\subsection{Statistical Tests of X-ray Variability}

To characterize variability, we performed different statistical tests on the data. First, we consider the coefficient of variation, defined as the ratio of the standard deviation to the mean, as follows:
\begin{equation}
\frac{\sigma}{\mu}=\frac{\sqrt{\sum_{i=1}^{n}(x_i-\mu)^2/(n-1)}}{\sum_{i=1}^{n}x_i/n},\label{eq_3}
\end{equation}
where $\sigma$ and $\mu$ are the standard deviation and the mean of data points, respectively, $i$ is the index of each data point $x_i$ in the time series, and $n$ is the total number of data points. This coefficient describes the extent of variations with respect to the mean, so its higher values are associated with stronger variability if we compare the same parameter. However, this coefficient is not suitable for comparison between dissimilar parameters and cannot evaluate the intrinsic nature of variations.
The ratio of the standard deviation of errors to that of data points ($\sigma_{\rm e}/\sigma$) may help us see whether the variability is intrinsic to the observation. Again, this ratio is suitable for evaluating values for the identical parameter. 
Both $\sigma/\mu$ and $\sigma_{\rm e}/\sigma$ are unable to characterize detailed features of consecutive series.
The normalized consecutive number, so-called \textit{Con}, which was first used by \citet{Wozniak2000}, may be used to quantify the changes at a number of consecutive points. To evaluate continuity in light curves, we define \textit{Con} 
as the number of three consecutive points greater or lower than $\sigma$ normalized by $n - 4$. For random fluctuations, \textit{Con} has a value of $\lesssim 0.045$ and is equal to 0 for constant consecutive series.

\begin{table*}
\begin{center}
\caption[]{Statistical results of timing analysis.
\label{rtcru:stat:result}}
\footnotesize
\begin{tabular}{clccccccccccc}
\hline\hline\noalign{\smallskip}
No.     &Param.     &$\sigma/\mu$        &$\sigma_{\rm e}/\sigma$     &$\eta$      &$\bar{\eta}_{\rm norm}$    &\multicolumn{2}{c}{Lilliefors test}   &\multicolumn{2}{c}{A--D test}   &\multicolumn{2}{c}{S--W test}   &\textit{Con}   \\
&     &        &     &      &    &$D$      &$p$-value   &$A^2$     &$p$-value   &$W $     &$p$-value   &   \\
\noalign{\smallskip}
\tableline
\noalign{\smallskip}
\multicolumn{13}{c}{\textit{XMM-Newton} (EPIC-pn)} \\
\noalign{\smallskip}
 &S+M+H        & $    0.382 $ & $    0.159 $ & $    1.138 $ & $    2.023 \pm  0.420 $ & $    0.083 $ & $    0.188 $ & $    0.827 $ & $    0.032 $ & $    0.968 $ & $    0.030 $ & $    0.106 $ \\
&HR$_1$        & $    1.234 $ & $    0.318 $ & $    2.242 $ & $    2.023 \pm   0.420 $ & $    0.054 $ & $    0.797 $ & $    0.251 $ & $    0.734 $ & $    0.993 $ & $    0.908 $ & $    0.059 $ \\
&HR$_2$        & $   -2.621 $ & $    0.247 $ & $    2.206 $ & $    2.023 \pm   0.420 $ & $    0.091 $ & $    0.098 $ & $    0.542 $ & $    0.159 $ & $    0.984 $ & $    0.348 $ & $    0.024 $ \\
\noalign{\smallskip}
\multicolumn{13}{c}{\textit{Suzaku} (XIS1)} \\
\noalign{\smallskip}
1&S+M+H        & $    0.153 $ & $    0.066 $ & $    0.881 $ & $    2.021 \pm   0.400 $ & $    0.068 $ & $    0.355 $ & $    0.501 $ & $    0.202 $ & $    0.982 $ & $    0.221 $ & $    0.064 $ \\
&HR$_1$        & $    0.223 $ & $    0.086 $ & $    1.159 $ & $    2.021 \pm   0.400 $ & $    0.095 $ & $    0.037 $ & $    0.731 $ & $    0.055 $ & $    0.979 $ & $    0.119 $ & $    0.043 $ \\
&HR$_2$        & $    0.306 $ & $    0.062 $ & $    0.591 $ & $    2.021 \pm   0.400 $ & $    0.079 $ & $    0.176 $ & $    0.539 $ & $    0.163 $ & $    0.984 $ & $    0.282 $ & $    0.128 $ \\
\noalign{\smallskip}
2&S+M+H        & $    0.206 $ & $    0.054 $ & $    1.257 $ & $    2.025 \pm   0.433 $ & $    0.077 $ & $    0.337 $ & $    0.265 $ & $    0.687 $ & $    0.991 $ & $    0.835 $ & $    0.050 $ \\
&HR$_1$        & $    0.497 $ & $    0.132 $ & $    1.621 $ & $    2.025 \pm   0.433 $ & $    0.104 $ & $    0.043 $ & $    0.897 $ & $    0.021 $ & $    0.964 $ & $    0.021 $ & $    0.000 $ \\
&HR$_2$        & $    0.118 $ & $    0.115 $ & $    1.398 $ & $    2.025 \pm   0.433 $ & $    0.075 $ & $    0.380 $ & $    0.321 $ & $    0.525 $ & $    0.987 $ & $    0.578 $ & $    0.013 $ \\
\noalign{\smallskip}
Mix&S+M+H        & $    0.277 $ & $    0.038 $ & $    0.470 $ & $    2.011 \pm   0.294 $ & $    0.067 $ & $    0.062 $ & $    1.075 $ & $    0.008 $ & $    0.982 $ & $    0.018 $ & $    0.142 $ \\
&HR$_1$        & $    0.539 $ & $    0.057 $ & $    0.385 $ & $    2.011 \pm   0.294 $ & $    0.109 $ & $    0.001 $ & $    2.724 $ & $    0.000 $ & $    0.966 $ & $    0.000 $ & $    0.068 $ \\
&HR$_2$        & $    0.313 $ & $    0.051 $ & $    0.337 $ & $    2.011 \pm   0.294 $ & $    0.090 $ & $    0.002 $ & $    1.532 $ & $    0.001 $ & $    0.973 $ & $    0.002 $ & $    0.193 $ \\
\noalign{\smallskip}
\multicolumn{13}{c}{\textit{Suzaku} (HXD-PIN)} \\
\noalign{\smallskip}
1&H+H$_{\rm E}$       & $    0.132 $ & $    0.094 $ & $    1.140 $ & $    2.038 \pm   0.533 $ & $    0.072 $ & $    0.707 $ & $    0.163 $ & $    0.941 $ & $    0.987 $ & $    0.814 $ & $    0.038 $ \\
&HR$_3$        & $    0.180 $ & $    0.210 $ & $    1.775 $ & $    2.038 \pm   0.533 $ & $    0.082 $ & $    0.503 $ & $    0.259 $ & $    0.703 $ & $    0.983 $ & $    0.617 $ & $    0.000 $ \\
&HR$_4$        & $   -0.272 $ & $    0.197 $ & $    1.898 $ & $    2.038 \pm   0.533 $ & $    0.082 $ & $    0.494 $ & $    0.249 $ & $    0.737 $ & $    0.985 $ & $    0.741 $ & $    0.000 $ \\
\noalign{\smallskip}
2&H+H$_{\rm E}$       & $    0.090 $ & $    0.161 $ & $    2.250 $ & $    2.067 \pm   0.704 $ & $    0.090 $ & $    0.746 $ & $    0.304 $ & $    0.551 $ & $    0.964 $ & $    0.378 $ & $    0.034 $ \\
&HR$_3$        & $    0.184 $ & $    0.154 $ & $    1.579 $ & $    2.067 \pm   0.704 $ & $    0.078 $ & $    0.902 $ & $    0.198 $ & $    0.878 $ & $    0.975 $ & $    0.675 $ & $    0.034 $ \\
&HR$_4$        & $   -0.707 $ & $    0.179 $ & $    1.542 $ & $    2.067 \pm   0.704 $ & $    0.097 $ & $    0.636 $ & $    0.294 $ & $    0.576 $ & $    0.970 $ & $    0.520 $ & $    0.034 $ \\
\noalign{\smallskip}
Mix&H+H$_{\rm E}$       & $    0.133 $ & $    0.099 $ & $    1.149 $ & $    2.024 \pm   0.425 $ & $    0.056 $ & $    0.776 $ & $    0.366 $ & $    0.427 $ & $    0.986 $ & $    0.483 $ & $    0.072 $ \\
&HR$_3$        & $    0.259 $ & $    0.129 $ & $    0.866 $ & $    2.024 \pm   0.425 $ & $    0.111 $ & $    0.018 $ & $    1.017 $ & $    0.011 $ & $    0.961 $ & $    0.011 $ & $    0.036 $ \\
&HR$_4$        & $   -0.504 $ & $    0.135 $ & $    0.910 $ & $    2.024 \pm   0.425 $ & $    0.093 $ & $    0.091 $ & $    0.511 $ & $    0.190 $ & $    0.984 $ & $    0.392 $ & $    0.048 $ \\
\noalign{\smallskip}
\multicolumn{13}{c}{\textit{NuSTAR} (FPMA+B)} \\
\noalign{\smallskip}
&H+H$_{\rm E}$       & $    0.145 $ & $    0.053 $ & $    1.046 $ & $    2.017 \pm   0.356 $ & $    0.096 $ & $    0.012 $ & $    2.196 $ & $    0.000 $ & $    0.908 $ & $    0.000 $ & $    0.050 $ \\
&HR$_3$        & $    0.108 $ & $    0.135 $ & $    1.898 $ & $    2.017 \pm   0.356 $ & $    0.089 $ & $    0.030 $ & $    1.706 $ & $    0.000 $ & $    0.845 $ & $    0.000 $ & $    0.042 $ \\
&HR$_4$        & $    5.704 $ & $    0.155 $ & $    1.932 $ & $    2.017 \pm   0.356 $ & $    0.059 $ & $    0.431 $ & $    0.747 $ & $    0.050 $ & $    0.930 $ & $    0.000 $ & $    0.025 $ \\
\noalign{\smallskip}\hline
\end{tabular}
\end{center}
\begin{tablenotes}
\footnotesize
\item[1]\textbf{Notes.} $\sigma/\mu$ is the ratio of the standard deviation to the mean, $\sigma_{\rm e}/\sigma$ is the ratio of the standard deviation of errors to that of data, $\eta$ is the von Neumann ratio, $\bar{\eta}_{\rm norm}$ the mean von Neumann ratio for no autocorrelation in the time series, and \textit{Con} is the normalized consecutive number. The hypothesis of normality can be evaluated using $p$-values from the Lilliefors, Anderson--Darling (A--D), and Shapiro--Wilk (S--W) statistical tests. The \textit{Suzaku} data labels 1, 2, and mix correspond to Obs. IDs 402040010, 906007010, and the combined observations, respectively.
\end{tablenotes}
\end{table*}

The von Neumann ratio \citep{vonNeumann1941}, which is the mean squared successive difference ($\delta^2$) with respect to the variance ($\sigma^2$), can also quantify the autocorrelation in successive series:
\begin{equation}
\eta=\frac{\delta^2}{\sigma^2}=\frac{\sum_{i=1}^{n-1}(x_{i+1}-x_i)^2/(n-1)}{\sigma^2}.\label{eq_4}
\end{equation}
For a normal distribution, the mean von Neumann ratio is expected to be $\bar{\eta}_{\rm norm}=2n/(n-1) \sim 2$ \citep{Young1941}. To assess the presence of autocorrelation, we estimated  the confidence intervals of $\bar{\eta}_{\rm norm}$ with the significance level of $\alpha=0.05$ for a normal probability distribution. A von Neumann ratio within the confidence levels of $\bar{\eta}_{\rm norm} \sim 2$ implies no autocorrelation, whereas values outside the aforementioned confidence levels toward 0 and 4 correspond to positive and negative autocorrelation in successive data points, respectively.

To examine the randomness of data points, we employ three non-parametric statistical tests of normality, namely the Lilliefors test, the Anderson--Darling test (hereafter referred to as the A--D test), and the Shapiro--Wilk test (hereafter referred to as the S--W test). The Lilliefors method \citep{Lilliefors1967} is a modification of the Kolmogorov--Smirnov (K--S) test, which utilizes estimated $\mu$ and $\sigma^2$ of the data for the assessment of normality. The K--S test is appropriate for the standard normal distribution $\mathcal{N}(\mu,\sigma^2)$ with $\mu=0$ and $\sigma^2=1$. The parameters for the Lilliefors method are based on the mean and variance of the specified data. The Lilliefors statistic determines the maximum difference ($D$) between the empirical distribution function (EDF) of the sample and the cumulative distribution function (CDF) of the normal distribution defined by the estimated mean and variance of the sample. The A--D test \citep{Anderson1952} is an extension of the Cram\'{e}r--von Mises statistic, which is based on the squared difference ($A^2$) between the EDF and the CDF with more weight to the tails of the distribution. The S--W test \citep{Shapiro1965} utilizes a statistical method ($W$) based on the order statistics, the expected values of the order statistics of independent and random variables from the normal distribution, and the covariance of the aforementioned order statistics. For the Lilliefors and A--D tests, we used the corresponding procedures from the \textsf{Statsmodels} package \citep{Seabold2010}, whereas the S--W test was performed with the relevant statistical function from the \textsf{SciPy} package \citep{Virtanen2020}. For all three tests, $p$-values less than or equal to the statistically significant level of $\alpha=0.05$ lead to the rejection of the hypothesis of normality, i.e., non-random variations.
The S--W test is the most powerful, followed closely by the A--D test and then the Lilliefors statistic, whereas the K--S method is less powerful than others \citep{Stephens1974,Razali2011}. 
However, the A--D test is more compelling than the S--W test in a population distribution with a very sharp peak and abruptly ended tails.

Table~\ref{rtcru:stat:result} summarizes the results of our statistical analysis of the X-ray variability in RT\,Cru obtained with different methods. It can be seen that $\sigma/\mu$ and $\sigma_{\rm e}/\sigma$ of $S+M+H$, HR$_{1}$, and HR$_{2}$, are higher in the \textit{XMM-Newton} data than those in the \textit{Suzaku} observations. This implies that variations are higher in the \textit{XMM-Newton} light curves, but with larger uncertainties. We also notice a higher $\sigma/\mu$ in the HR$_{4}$ ratio of the \textit{NuSTAR} data, which may be an indication of scattered fluctuations in this ratio.
However, $\sigma/\mu$ is not a suitable tool for quantifying the variability in a time sequence of data. The \textit{Con} number may be able to better distinguish variations in a consecutive sequence. The values of \textit{Con} indicate consecutive changes in $S+M+H$ in the \textit{XMM-Newton} and combined \textit{Suzaku}/XIS1 observations,  HR$_{2}$ in the first \textit{Suzaku}/XIS1 observation and two combined \textit{Suzaku}/XIS1 observations, while others with \textit{Con} $\lesssim 0.045$ may have random variations. Nevertheless, our \textit{Con} statistical method is based only on three consecutive data points, so it is unable to obtain a broader picture of variability in the entire sequence of data points. 

The von Neumann ratio ($\eta$) can effectively quantize the systematic structure of time series. Furthermore, the A--D and S--W tests of normality can properly determine whether or not a normal (random) distribution describes the variables. For the \textit{XMM-Newton} data, $\eta$ is not within the confidence range of $\bar{\eta}_{\rm norm} \sim 2$ in the $S+M+H$ broad band, while the $p$-values of  the A--D and S--W tests are below the significant level of $\alpha=0.05$, resulting in the rejection of the hypothesis of normality. However, the von Neumann ratios of HR$_{1}$ and HR$_{2}$ are within the range of $\bar{\eta}_{\rm norm}$, along with $p$-values of a normal distribution. Although the broad-band light curve of the \textit{XMM-Newton} data exhibits abnormal variability, there are no changes in the hardness conditions.

In the case of \textit{Suzaku}/XIS1, the von Neumann ratio demonstrates positive autocorrelation in the time series of the broad band ($S+M+H$) and hardness ratios of the first, second, and combined observations, apart from HR$_{1}$ in the second observation. However, the $p$-values of the normality tests suggest that there are not normal distributions in HR$_{1}$ of the first XIS1 observation with marginally significant statistics and the second XIS1 observation with significant statistics. Furthermore, the A--D and S--W tests of $S+M+H$, HR$_{1}$, and HR$_{2}$ in the two mixed \textit{Suzaku}/XIS1 observations definitely reject the hypothesis of normality, which is obvious in the movement pattern seen in the hardness diagrams in Fig.~\ref{rtcru:fig:hdr}. Nevertheless, we see a non-normal distribution only in the HR$_{3}$ ratio of the \textit{Suzaku}/PIN observations, but not $H+H_{E}$ and HR$_{4}$. The von Neumann statistical analysis of the \textit{Suzaku}/PIN data also depicts positive autocorrelation on the $H+H_{E}$ broad band of the first and combined observations, and on HR$_{3}$ and HR$_{4}$ of the mixed multi-epoch data, which are consistent with the spectral transition occurring between 2007 and 2012.

The broad band ($H+H_{E}$) and the hardness ratios (HR$_{3}$ and HR$_{4}$) of the \textit{NuSTAR} data exhibit non-random distributions according to the $p$-values obtained from A--D and S--W tests. However, von Neumann's mean squared successive difference depicts a positive autocorrelation only in the broad band. The hardness diagrams in Fig.~\ref{rtcru:fig:hdr} seem to not depict any obvious hardness transition in the \textit{NuSTAR} data, though our A--D and S--W normality tests apparently suggest it, which is not supported by the von Neumann statistics. However, all of the normality tests and the von Neumann ratio indicate that there are statistically significant fluctuations in the \textit{NuSTAR} broad band.

In summary, our statistical analysis of the variability shows that RT\,Cru underwent flux variations over the broad band of the \textit{XMM-Newton} data based on $\eta$ and the A--D/S--W test, 
the first and second \textit{Suzaku}/XIS1 data, the first and combined Suzaku/PIN data, the \textit{NuSTAR} data according to the von Neumann results, and the mixed first-second \textit{Suzaku}/XIS1 data and the \textit{NuSTAR} data based on the A--D/S--W test.
Moreover, our results suggest that the X-ray source experienced a long-term spectral transition between the first and second \textit{Suzaku} observations based on the normality tests of HR$_{1}$, HR$_{2}$, and HR$_{3}$, as well as some short-term hardness fluctuations in HR$_{1}$ of the second \textit{Suzaku}/XIS1 data and HR$_{3}$ of the \textit{NuSTAR} data.

\section{Principal Component Analysis}
\label{rtcru:pca}

\subsection{Computational Approach}
\label{rtcru:pca:method}

Principal Components Analysis (PCA), which is a highly effective technique for deconstructing temporally variable data, has frequently been used in the field of multivariate statistical research in astronomy, such as gaining insights into the spectral types of stars \citep{Deeming1964,Whitney1983}, statistical analysis of galaxies \citep{Bujarrabal1981,Efstathiou1984}, multiple-epoch UV data of active galaxies \citep{Mittaz1990}, distinct emission components in optical observations of quasars \citep{Francis1992,Boroson1992}, and imaging analysis of molecular clouds \citep{Heyer1997,Brunt2002a}. 
This eigenvector-based multivariate method was later employed to analyze the X-ray spectral variability of Seyfert galaxies
\citep{Vaughan2004,Miller2007,Parker2014,Gallo2015} and X-ray binaries \citep{Malzac2006,Koljonen2013,Koljonen2015}.
The PCA mathematical framework for spectral analysis is similar to that adopted in the flux--flux correlation method as shown by \citet{Vaughan2004}, which results in multiple spectrally invariant components whose amplitudes exhibit temporal variation. The observed variable spectra can be regenerated by linearly combining all these PCA components. In theory, this approach should enable us to identify all the variable components of any given multidimensional dataset. However, the assumption of spectrally invariant components may not hold true in real observational data, where noise in the background is likely to cause some variations. As demonstrated by \citet{Parker2015}, one of the key benefits of employing multivariate analysis is the ability to determine the minimum number of PCA components required for building the observed variations, which may not be created by noise fluctuations.
A clear result of PCA would be time-dependent emission or absorption components with little variability but enough statistical significance to not be distinguished as background noise 
\citep[see, e.g.,][]{Koljonen2013,Parker2015,Parker2017}.

To conduct our principal component analysis of RT\,Cru, we used a customized implementation of the Python program \textsc{pca}  originally written by \citet{Parker2018}. This program leveraged the singular value decomposition (\textsc{svd}) function \citep[][]{Press1997} from the linear algebra (\textsf{linalg}) submodule of the Python library \textsf{NumPy}. It also utilizes the effective area column read from the ARF data to convert the count spectrum $C(E)$ to the photon-flux one $F_{\rm ph}(E)$, i.e., $F_{\rm ph}(E)= C(E) / (A_{\rm eff}(E) t_{\rm exp})$, where $A_{\rm eff}(E)$ is the effective function and $t_{\rm exp}$ the exposure time. The time-sliced spectrum of the background is also used to eliminate background contamination from the source. Moreover, a mean spectrum, $F_{\rm ph,m}(E)$ calculated with the time-sliced spectra of each dataset was used to derive the normalized time-sliced spectra as follows: $F_{{\rm n},k}(E) = (F_{{\rm ph},k}(E) - F_{\rm ph,m}(E))/F_{\rm ph,m}(E)$. The purpose of this program is to transfer a set of $n_t$ time-sliced spectra binned at $n_E$ energy intervals into a 2D array ($n_E \times n_t$). The resulting array is then subjected to a decomposition process using the \textsc{svd} function, resulting in a matrix ($n_E \times n_E$) including the principal components $f_{n_E} (E)$ sorted with their eigenvalues, in addition to an $n_t$-array of eigenvalues yielding the fractional variability, as well as a matrix ($n_t \times n_t$) containing the eigenvectors representing the time series $A_{n_t}(t)$. The PCA spectra and their corresponding light curves describe the spectral characteristics and their temporal variations present in a complex variable source, respectively. The fractional variability of each component is estimated using the normalized eigenvalues. The uncertainty calculation in the PCA program is implemented according to the procedure outlined in \citet{Miller2007}, where the spectra are subjected to random perturbations and subsequent recalculations with the \textsc{svd} function. 

\subsection{PCA Results}
\label{rtcru:pca:result}

\begin{figure*}
\begin{center}
\includegraphics[width=0.65\textwidth, trim = 0 0 0 0, clip, angle=0]{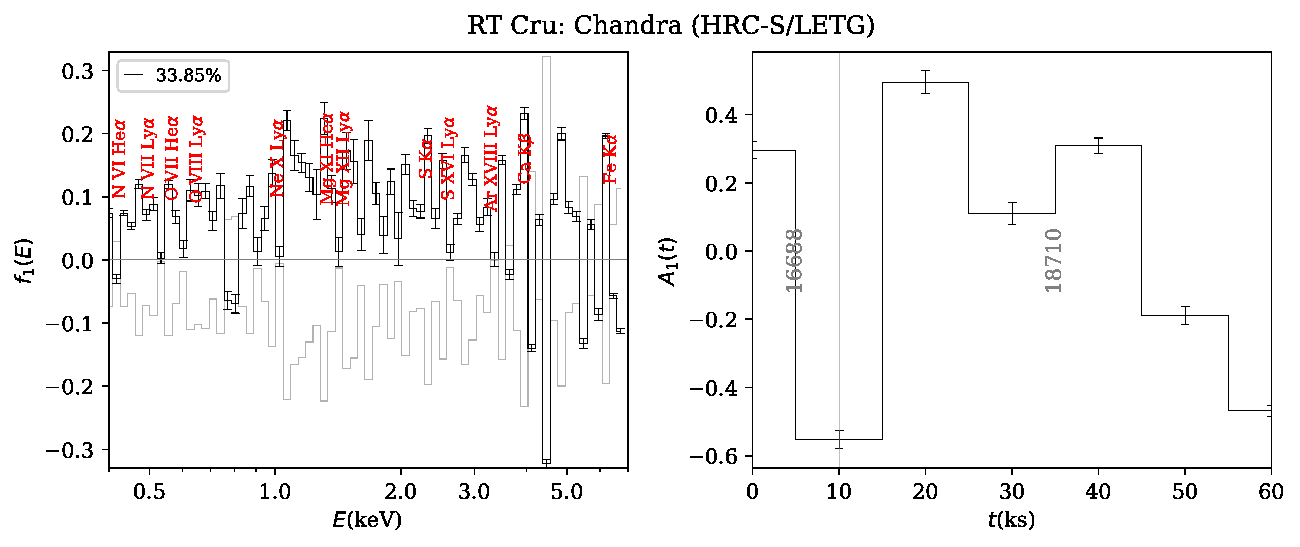}%
\includegraphics[width=0.3\textwidth, trim = 0 0 0 0, clip, angle=0]{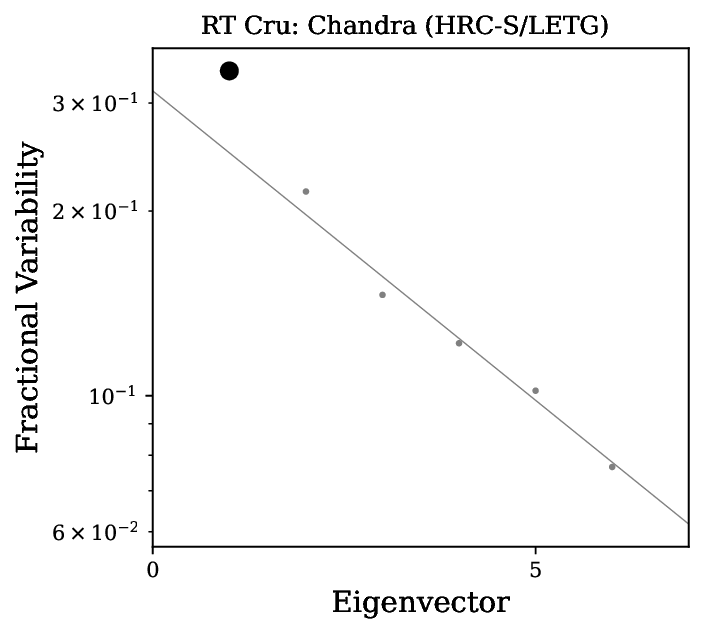}\\
\includegraphics[width=0.65\textwidth, trim = 0 0 0 0, clip, angle=0]{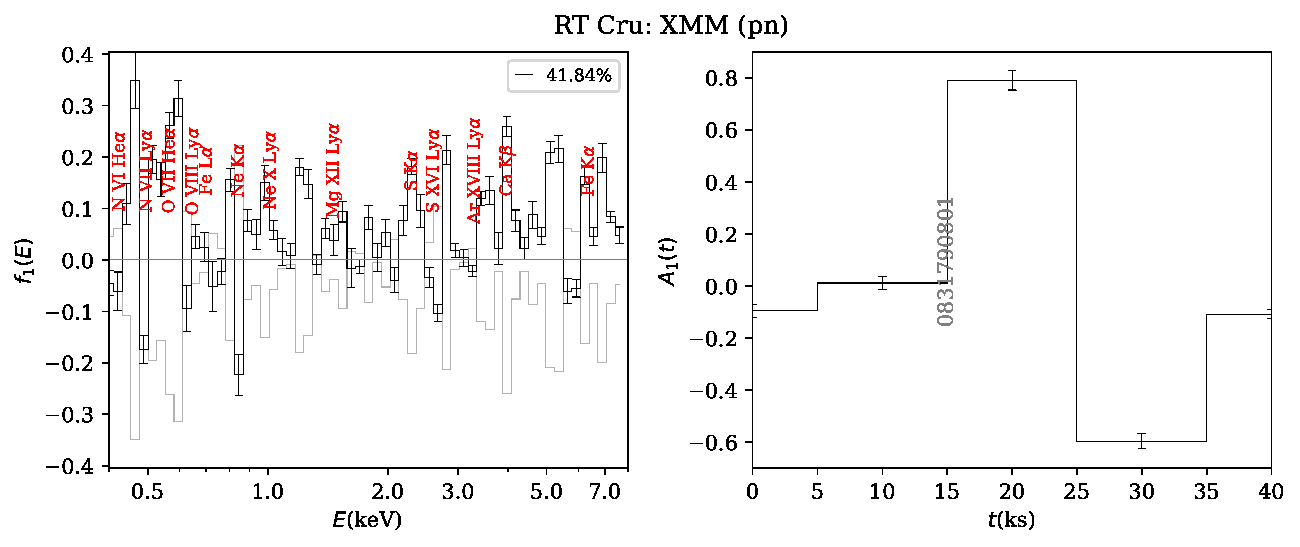}%
\includegraphics[width=0.3\textwidth, trim = 0 0 0 0, clip, angle=0]{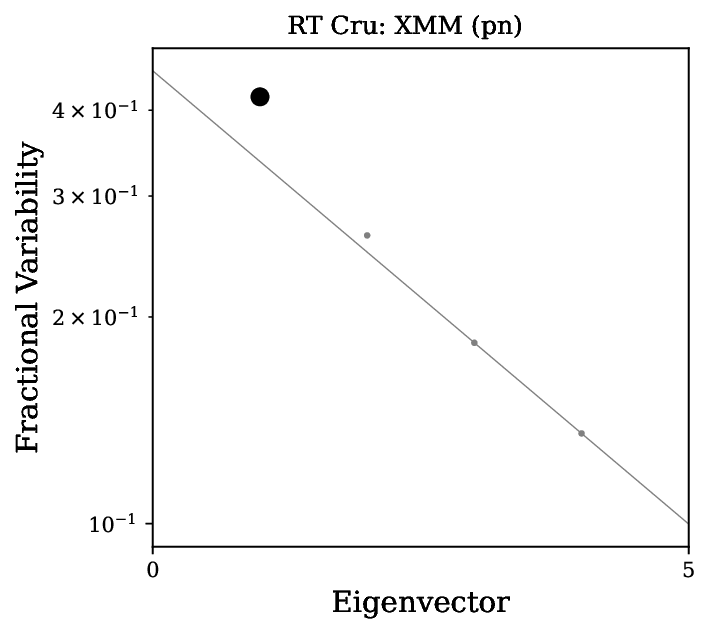}\\
\includegraphics[width=0.65\textwidth, trim = 0 0 0 0, clip, angle=0]{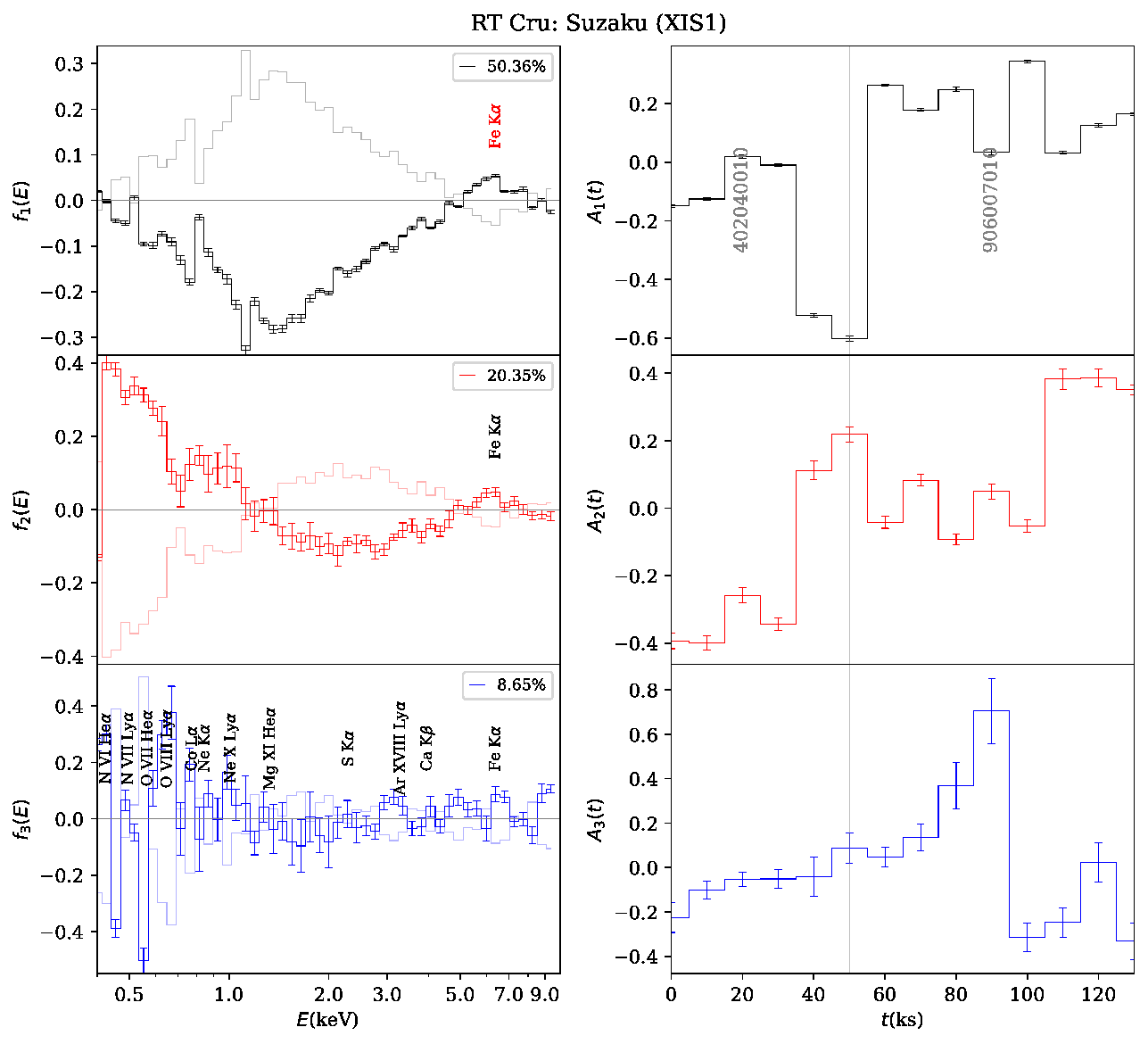}%
\includegraphics[width=0.3\textwidth, trim = 0 0 0 0, clip, angle=0]{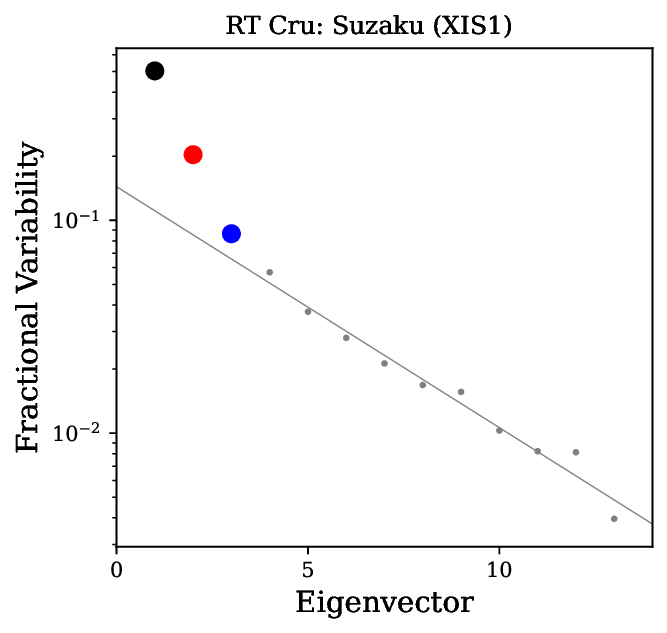}
\end{center}
\caption{The normalized spectrum $f_{i}(E)$ (left) and the corresponding time series $A_{i}(t)$ (middle) of the $i$th-order principal component determined from the \textit{Chandra} HRC-S/LETG (top panels), \textit{XMM-Newton} EPIC-pn data (middle panels), and \textit{Suzaku} XIS1 data (bottom panels) of RT\,Cru, along with the LEV diagram (right) showing the linear correlation between logarithmic normalized eigenvalues and eigenvector orders among the higher-order ($>$\,3) components. The energy levels, where thermal emission lines may be present, are marked in the spectra.
The number of non-zero eigenvalues in each LEV diagram is associated with the number of time-segmented spectra.
\label{rtcru:fig:pca:1}
}
\end{figure*}

Figure~\ref{rtcru:fig:pca:1} shows the spectrum and its time series of the first components produced from the \textit{Chandra} HRC-S/LETG (top panels) and \textit{XMM-Newton} EPIC-pn data (middle panels), along with the corresponding log-eigenvalue (LEV) diagrams (right). We should note that the number of non-zero eigenvalues provided by the \textsc{svd} function corresponds to the number ($n_t$) of time-segmented pulse-height amplitude (PHA) spectra in each LEV diagram. The black line in each LEV diagram corresponds to a linear correlation found between logarithmic normalized eigenvalues and eigenvector orders of the components with orders higher than 3, which should be associated with noise rather than real spectral variations. The LEV diagrams help us determine those components that are statistically significant as described by \citet[][]{Parker2018}. It can be seen that the first component in each dataset, whose spectrum and time series are plotted, is slightly above the high-order correlation line.
The peak and valley features that appear in the PCA spectra could be indicators of emission and/or absorption lines \citep[see, e.g.,][]{Parker2017,Parker2018}. We see some spectral features in the soft band, which might suggest the presence of emission lines from H-like and He-like ions, \ionic{N}{vii} Ly$\alpha$ 0.5\,keV, \ionic{N}{vi} He$\alpha$ 0.43\,keV, \ionic{O}{viii} Ly$\alpha$ 0.65\,keV, and \ionic{O}{vii} He$\alpha$ 0.57\,keV. 
However, it is difficult to detect these thermal emission features in the soft excess, even if they do exist, because they are mixed up with high background noise. A recent investigation by \citet{Zhang2023} could not constrain these emission lines but put an upper confidence level of 1\,keV on the temperature of a soft thermal plasma component. The time series $A_{1}(t)$ of the first components suggest that these emission lines, even if they are real, appear temporarily during brightening events that occur every 20-25\,ks and last for $\lesssim 10$\,ks, which makes it difficult to constrain them. Although the effectiveness of the \textit{Chandra} HRC-S/LETG in the hard band is low, the \textit{XMM-Newton} EPIC-pn instrument seems to capture some weak features in the energies associated with \ionic{Ar}{xviii} Ly$\alpha$ 3.3\,keV, Ca K$\beta$ 4\,keV, and Fe K$\alpha$ 6.4\,keV. Nevertheless, the emission lines look to be more predominant in the soft band ($<1$\,keV) than the hard band in the $f_{1}(E)$ spectrum of the \textit{XMM-Newton} data.

In Fig.~\ref{rtcru:fig:pca:1}, we also present the spectra $f_{i}(E)$ and the corresponding light curves $A_{i}(t)$ of the first three PCA components generated from the \textit{Suzaku} XIS1 data. These three components are statistically significant according to their noticeable deviations from the high-order eigenvalue regression in the LEV diagram. As seen in Fig.\,\ref{rtcru:fig:hdr}, the source exhibited increases in both the softness (lower $\mathrm{HR}_{1}$) and the hardness (higher $\mathrm{HR}_{2}$) in 2012 compared to 2007, while the brightness in the total band ($S+M+H$) was lower in 2012 than in 2007.

The first component in the \textit{Suzaku} XIS1 data is akin to the spectral features of an absorbing column component.
The second component shows a blackbody-like emission below 2\,keV followed by an absorbed hard continuum above 2\,keV, but with a peak at $\sim6.4$\,keV, corresponding to the iron K$\alpha$ line. The first two components depict a highly absorbed continuum source, which is similar to the model used by \citet[][]{Danehkar2021}. Additionally, our multivariate analysis suggests a (multi-) blackbody-like thermal component in the soft excess in the \textit{Suzaku} observations. 
The normalized eigenvalues yield the variability fractions of $\sim 50$\% and 20\% for the spectra $f_{1}(E)$ and $f_{2}(E)$, respectively.
This implies that the variabilities in the absorbing material and the source continuum are mainly responsible for the changes in the hardness ratios $\mathrm{HR}_{1}$ and $\mathrm{HR}_{2}$ between 2007 and 2012.
This phenomenon might be related to the almost total disappearance of the hard X-rays identified in 2019, which was argued to be associated with a substantial reduction in the falling material \citep{Pujol2023}. In addition, the absorbing column $f_{1}(E)$ contributes about twice the dramatic changes in the hardness ratios of the source continuum $f_{2}(E)$ over the two epochs. The light curves $A_{1}(t)$ and $A_{2}(t)$ depict that as the source got fainter and harder in the last epoch than in the previous epoch, it had higher obscuration in 2012 than in 2007.

\begin{figure*}
\begin{center}
\includegraphics[width=0.65\textwidth, trim = 0 0 0 0, clip, angle=0]{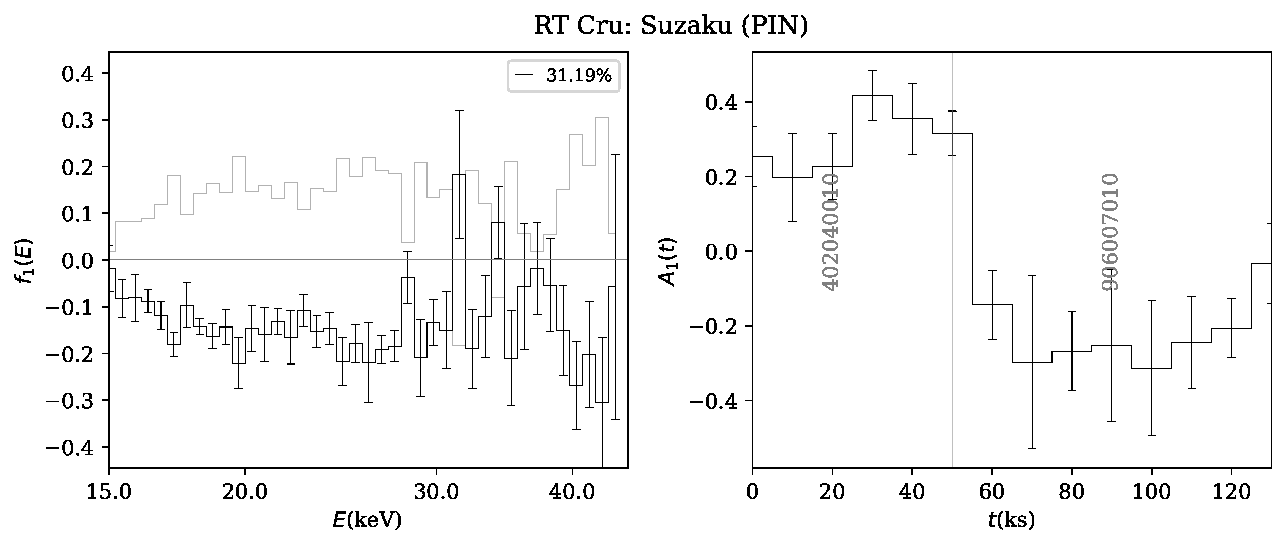}%
\includegraphics[width=0.3\textwidth, trim = 0 0 0 0, clip, angle=0]{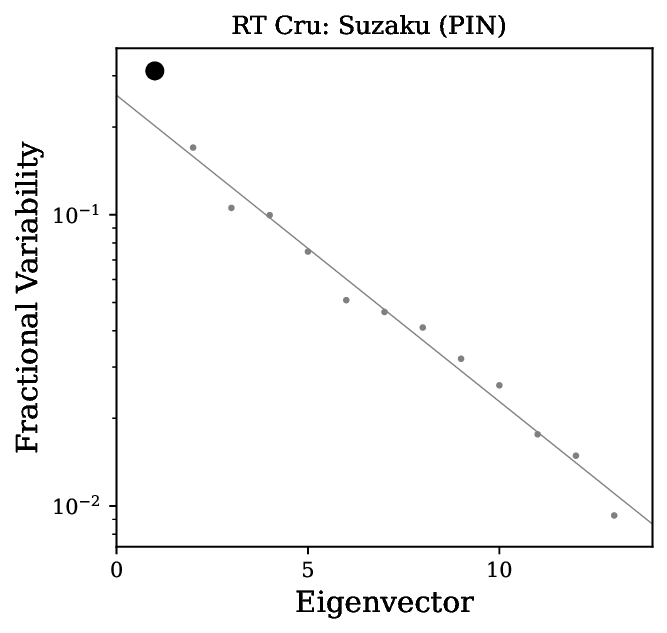}
\includegraphics[width=0.65\textwidth, trim = 0 0 0 0, clip, angle=0]{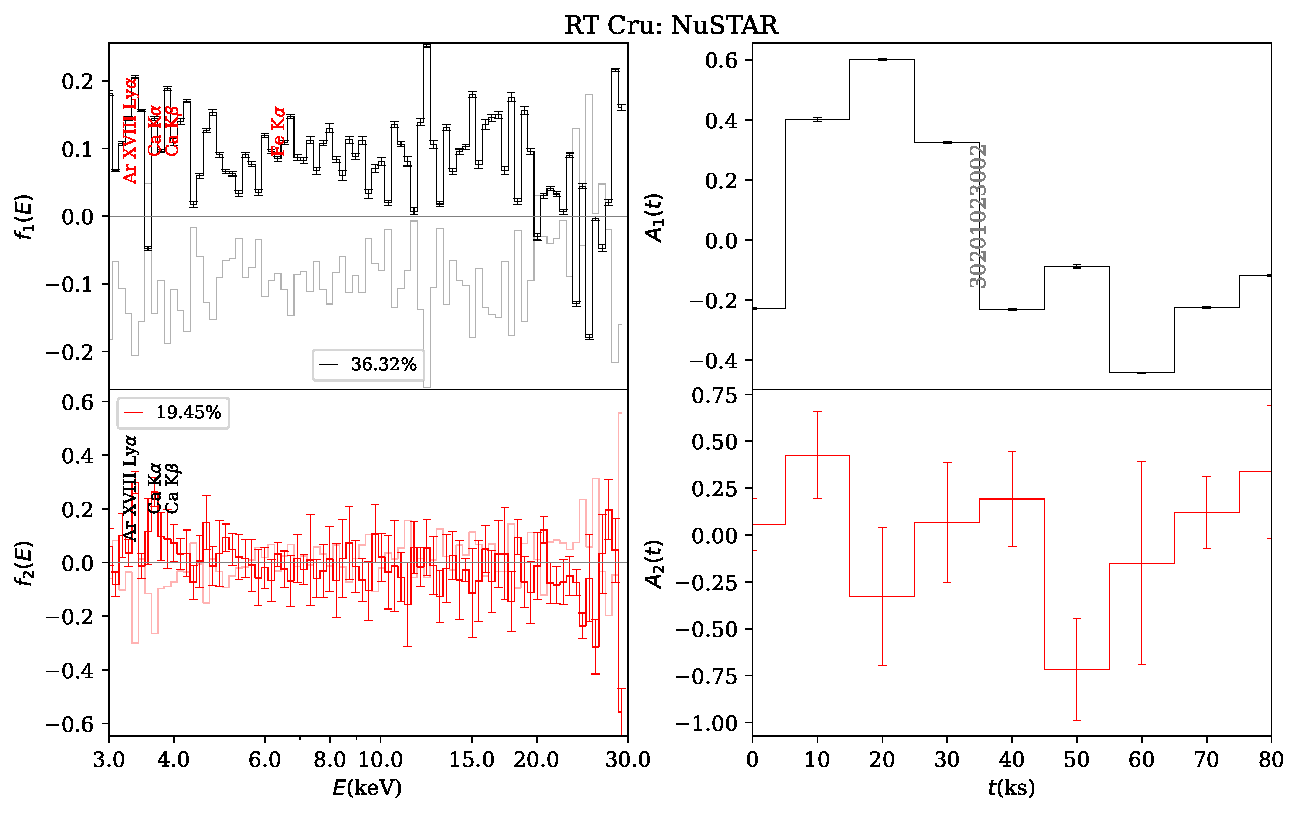}%
\includegraphics[width=0.3\textwidth, trim = 0 0 0 0, clip, angle=0]{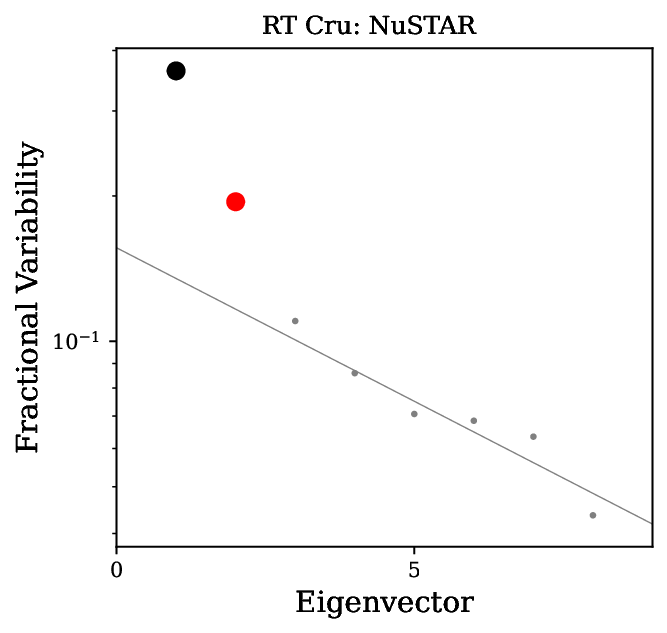}\\
\end{center}
\caption{The same as Fig.~\ref{rtcru:fig:pca:1}, but for the \textit{Suzaku} HXD-PIN (top panels) and \textit{NuSTAR} data (bottom panels) of RT\,Cru.
\label{rtcru:fig:pca:2}
}
\end{figure*}

The third-order PCA component derived from the \textit{Suzaku} XIS1, which has a small variability fraction of $\sim 9$\%, has some spectral features similar to those in the first-order PCA spectrum deduced from the \textit{XMM-Newton} EPIC-pn data. Once again, the emission features in the soft excess ($<1$\,keV) are stronger than those in the hard excess. The time series $A_{3}(t)$ also indicates that these emission lines, while potentially of physical origin, are likely related to the flickering nature. They apparently originate from a brightening event that started 15\,ks after the beginning of the second observation and lasted for $\sim 20$\,ks.

The spectrum and its light curve of the first PCA component extracted from the \textit{Suzaku} HXD-PIN data are shown in Fig.~\ref{rtcru:fig:pca:2}, along with the log-eigenvalue diagram. The comparison between the time series of the XIS1 and PIN data indicates that both of them have similar changes over time, so the PCA spectrum $f_{1}(E)$ derived from the PIN observations corresponds to the extreme hard excess of the continuum seen in $f_{2}(E)$ of the XIS1 data. However, there is no indication of the absorbing column in the eigenvector-based multivariate analysis of the PIN data. This could mean that the absorbing material mostly blocks the energy range below 5\,keV, as seen in $f_{1}(E)$ of the XIS1 in Fig.~\ref{rtcru:fig:pca:1}. In addition, there is no PCA component containing the thermal emission lines similar to the spectrum $f_{3}(E)$ of the XIS1, so they could be the features predominately present in the soft excess. We should also note that the PIN has an energy resolution of around 3\,keV in the 10--30\,keV range, which is lower than the \textit{Suzaku} XIS (120\,eV at 6\,keV), so the PIN cannot resolve any emission line features.

Figure~\ref{rtcru:fig:pca:2} also presents the spectra $f_{i}(E)$ and the associated time series $A_{i}(t)$ of the first two PCA components deduced from the \textit{NuSTAR} observation. The first-order PCA spectrum $f_{1}(E)$ shows a powerlaw-like continuum similar to that seen in $f_{1}(E)$ of the \textit{Suzaku} HXD-PIN observations. Additionally, the second PCA component may contain some emission features at energies typically associated with \ionic{Ar}{xviii} Ly$\alpha$ 3.3\,keV, Ca K$\alpha$ 3.7\,keV, and Ca K$\beta$ 4\,keV, which are possibly present in $f_{1}(E)$ of the XMM-Newton EPIC-pn data and $f_{3}(E)$ of the \textit{Suzaku} XIS1 observations. 
The light curve $A_{2}(t)$ of the \textit{NuSTAR} PCA suggests that the second component, which likely contains some line features, appears during brightening events occurring at intervals between 30 and 40\,ks.
However, we caution that \textit{NuSTAR}'s energy resolution of $400$\,eV is lower than those of the \textit{XMM-Newon} EPIC-pn (80\,eV) and \textit{Suzaku} XIS (50\,eV at 1\,keV). 
In addition, there is no evidence for the obscuring material in the multivariate statistical analysis of the \textit{NuSTAR} data, implying the absorbing material is not largely variable during the course of the \textit{NuSTAR} observation similar to that seen in the \textit{Suzaku} observations. 

\subsection{Constructing Spectra from PCA}
\label{rtcru:pca:construct}

To better evaluate the nature of the PCA components from \textit{Suzaku}/XIS1, we used their corresponding photon-flux data to create the \textsc{xspec}-compatible spectra and the response files using the \textsc{ftool} program \textsf{ftflx2xsp}. The photon-flux spectra were reconstructed according to $F_{{\rm ph},i}(E)  = \lambda_{i}  (F_{{\rm n},i}(E) F_{\rm ph,m}(E)+F_{\rm ph,m}(E))$, where $F_{{\rm n},i}(E)$ is the normalized spectrum of the $i$th-order principal component produced by the \textsc{svd} function, $F_{\rm ph,m}(E)$ is the mean spectrum derived from the time-sliced spectra, and $\lambda_{i}$ is the normalized eigenvalue of the $i$th-order principal component created via \textsc{svd}, representing the variability fractions. We analyzed the reconstructed spectra in the Interactive Spectral Interpretation System \citep[\textsc{isis} v\,1.6.2-51;][]{Houck2000} that has access to the \textsc{xspec} models \citep[][]{Arnaud1996}.

\begin{figure}
\begin{center}
\includegraphics[width=0.45\textwidth, trim = 0 0 0 0, clip, angle=0]{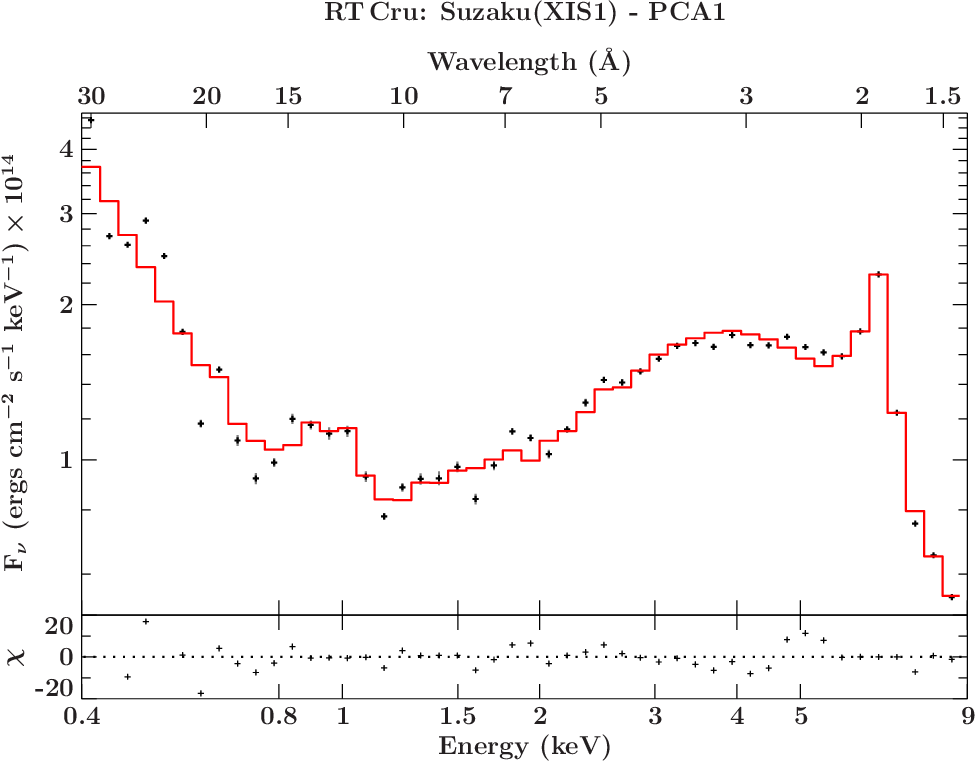}\\\medskip
\includegraphics[width=0.45\textwidth, trim = 0 0 0 0, clip, angle=0]{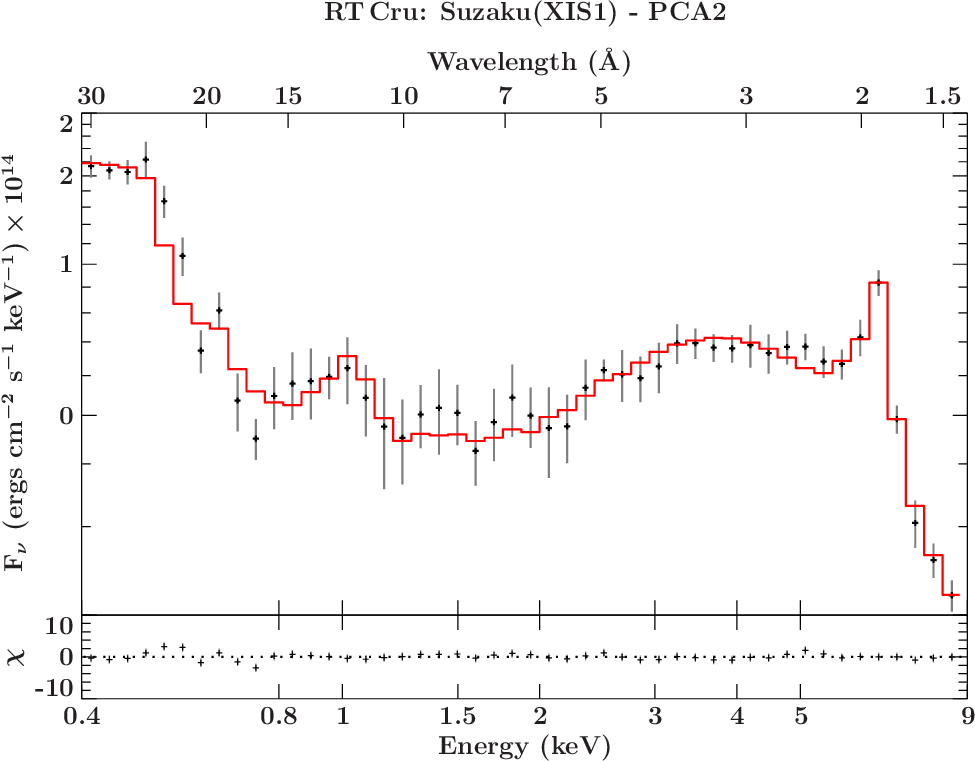}
\end{center}
\caption{The spectra of the first (top) and second (bottom) PCA components derived from the \textit{Suzaku} XIS1 data of RT\,Cru (black color), along with the fitted phenomenological model \textsf{pcfabs}\,$\times$\,(\textsf{diskbb}\,$+$\,\textsf{apec}) $+$ \textsf{pcfabs}\,$\times$\,(\textsf{compTT}\,$+$\,$\sum$\,\textsf{zgauss}) (red color).
\label{rtcru:fig:pca:model}
}
\end{figure}

Figure~\ref{rtcru:fig:pca:model} shows the reconstructed spectra of the first and second principal components of the \textit{Suzaku} XIS1 data, which were modeled using a phenomenological model, \textsf{pcfabs}\,$\times$\,(\textsf{diskbb} $+$ \textsf{apec}) $+$ \textsf{pcfabs}\,$\times$\,(\textsf{compTT} $+$ $\sum$\,\textsf{zgauss}), consisting of an accretion disk model (\textsf{diskbb}), a collisionally ionized diffuse model of the Astrophysical Plasma Emission Code \citep[\textsc{apec};][]{Smith2001a}, a Comptonization model (\textsf{compTT}), partial covering fraction absorption components (\textsf{pcfabs}), and Gaussian components (\textsf{zgauss}). The soft excess was well reproduced using a \textsf{diskbb} model made of multiple blackbody components \citep[see, e.g.,][]{Mitsuda1984,Makishima1986} and an emission spectrum produced by the collisional plasma \textsc{apec} model; both of them are partially covered by an absorbing column. To create phenomenologically the curvature in the hard excess, we employed a partially covered, absorbed Comptonization model (\textsf{compTT}) of the soft radiation in a hot plasma cloud analytically obtained by \citet{Titarchuk1994} and \citet{Titarchuk1995}. To improve the model fit, three Gaussian components were also included, namely Fe K$\alpha$ (6.379\,keV), Fe He$\alpha$ (6.693\,keV), and Fe Ly$\alpha$ (6.946\,keV). We caution that the PCA components are built from the \textit{Suzaku} data according to the temporal evolution over the two epochs, so not all the derived spectral features are physically real.

Table~\ref{rtcru:model:param} lists the best-fitting values of the parameters, obtained using the Levenberg--Marquardt minimization algorithm \citep[][]{More1978} and the chi-square statistic \citep[$\chi^2$; see][]{Bevington2003}, with the uncertainties at 90\% confidence derived using the \textsc{isis} standard function for confidence limits (\textsf{conf\_loop}).
It can be seen that the soft excess is partially covered by an absorbing column of $\sim 6 \times 10^{22}$\,cm$^{-2}$.
However, the covering fraction in the first PCA spectrum ($C_{f,{\rm dsk}}=0.95$) is higher than that in the second PCA spectrum ($C_{f,{\rm dsk}}=0.67$). Moreover, the hard excess is partially absorbed by columns of $\sim 1.7\times 10^{22}$ and $0.2 \times 10^{22}$\,cm$^{-2}$ in the first and second PCA components, respectively; both of them with a covering fraction of $C_{f,{\rm cmp}}=0.95$. This implies that there is more absorbing material in the first principal component. Interestingly, \cite{Luna2018} also found that the absorbing column in 2012 is about $3.4 \times 10^{22}$\,cm$^{-2}$ higher than that in 2007 based on the XIS1 data.

In addition, the curvature in the soft excess was phenomenologically reproduced using a multiple-blackbody accretion disk with a temperature of $\sim 1.8$\,keV at the inner radius, along with a collisionally ionized soft thermal plasma emission with temperatures of 1.1 and 1.2\,keV in the first and second PCA components, respectively. 
We should note that the inclusion of the \textsf{apec} collisional plasma model helps to recreate the PCA spectra, while it could not be done using only a \textsf{diskbb} model. In particular, \citet{Danehkar2021} previously deduced a heavily obscured, soft thermal plasma emission with a temperature of 1.3\,keV using low-count Bayesian statistics. 
Moreover, the third PCA component of the \textit{Suzaku} XIS1 data, along with the principal component of the \textit{XMM-Newton} EPIC-pn data, suggests the possible presence of some emission features mostly at the soft excess, which could be associated with thermal features of a collisionally ionized plasma. So, the soft thermal \textsf{apec} component of $\sim 1.1$--$1.2$\,keV  that appears in the spectral analysis of the PCA components might be physically related to that phenomenon. 

The soft component with a temperature of $\sim 1.2$\,keV, if it exists, could be produced by the expanding winds or jets, similar to what was proposed for CH\,Cyg \citep{Ezuka1998}. The predicted X-ray temperature of the shock-ionized plasma created by the collision of the winds with the interstellar medium is $kT=(3/16)\mu m_{\rm H}v^2$, where $v$ is the wind velocity, $\mu=0.615$ is the mean molecular weight, and $m_{\rm H}$ the hydrogen mass \citep[e.g.,][]{Guedel2009}. To produce an X-ray thermal spectrum of 1.2\,keV, a wind velocity of 1000 km\,s$^{-1}$ is necessary. This is in the range of those estimated for CH\,Cyg \citep{Karovska2007,Karovska2010}, which is another $\delta$-type symbiotic star containing soft thermal spectra of 0.2 and 0.7\,keV \citep[][]{Ezuka1998}.

\begin{table}
\begin{center}
\caption{Best-fitting parameters for the phenomenological model of the first and second PCA components of the \textit{Suzaku} XIS1 data.  
\label{rtcru:model:param}
}
\begin{tabular}{lccc}
\hline\hline
\noalign{\smallskip}
{\textsc{xspec}} & {Parameter} & {PCA1}   & {PCA2}   \\
\noalign{\smallskip}
\hline 
\noalign{\smallskip}
\textsf{pcfabs} & $N_{\rm H,dsk}(10^{22}\mathrm{cm}^{-2})$ \dotfill & $ {  6.65 }_{ -0.10 }^{ +0.10 } $ & $ {  6.40 }_{ -1.08 }^{  +1.26 } $   \\ 
                & $C_{f,{\rm dsk}}$ \dotfill & $ {  0.95 }_{ -0.001}^{  +0.001} $ & $ {  0.67 }_{ -0.03 }^{  +0.03 } $      \\
\noalign{\smallskip}
\textsf{diskbb}  & $T_{\rm in}$ (keV) \dotfill  &$ {  1.76 }_{ -0.01 }^{  +0.01 } $   & $ {  1.80 }_{ -0.07 }^{  +0.07 } $  \\  
                 & $K_{\rm dsk}$ ($10^{-5}$)  \dotfill & $91.82 $    &  $ 32.94 $ \\
\noalign{\smallskip}
\textsf{apec}   & $kT$ (keV) \dotfill &  $ {  1.05 }_{ -0.01 }^{  +0.01 } $ & $ {  1.22 }_{ -0.11 }^{  +0.14 } $ \\  
                 & $K_{\rm apc}$ ($10^{-5}$)  \dotfill & $ 1.85 $    &  $ 0.17 $ \\
\noalign{\medskip}  
\textsf{pcfabs} & $N_{\rm H,cmp}(10^{22}\mathrm{cm}^{-2})$ \dotfill & $ {  1.68 }_{ -0.03 }^{  +0.02 } $ & $ {  0.18 }_{ -0.05 }^{  +0.05 } $   \\ 
                & $C_{f,{\rm cmp}}$ \dotfill & $ {  0.95 }_{ -0.001}^{  +0.001} $ & $ {  0.95 }_{ -0.03 }^{  +0.001 } $      \\
\noalign{\smallskip}                 
\textsf{compTT} & $T_{0}$ (keV) \dotfill  & $ {  0.03 }_{ -0.001}^{  +0.001} $   &   $ {  0.06 }_{ -0.001 }^{  +0.001} $ \\ 
                   & $kT$ (keV) \dotfill & $ { 94.24 }_{ -36.86 }^{  +1.65 } $    &  $ {  2.00 }_{   }^{  +4.87 } $ \\
                   & $\tau$ \dotfill & $ 0.01 $    &  $ 0.01 $ \\
                   & $K_{\rm cmp}$ ($10^{-5}$)  \dotfill & $ 6.16 $    &  $ 11.66 $ \\
\noalign{\smallskip}
\hline
\noalign{\smallskip}
\end{tabular}
\end{center}
\end{table}

\vfill\break

\subsection{Simulating Variability}
\label{rtcru:pca:sim}

To trace the origin of the X-ray variability, we simulated X-ray spectra using the \textsc{isis} function \textsf{fakeit} with the corresponding response data of the \textit{XMM-Newton}/EPIC-pn and \textit{Suzaku}/XIS1 observations. We implemented our simulations in a manner similar to \citet{Koljonen2013} and \citet{Parker2014a}, but instead of using random changes between the confidence limits as they did, we reproduced the X-ray variability using the PCA time series extracted from the observations. Sets of the simulated time-sliced spectra with an exposure interval of 10\,ks were created and stored into PHA files with the aid of the Remeis \textsc{isis} functions (ISISscripts). We loaded and explored them with the same PCA program used for our analysis. 

\begin{figure}
\begin{center}
\includegraphics[width=0.45\textwidth, trim = 0 0 0 0, clip, angle=0]{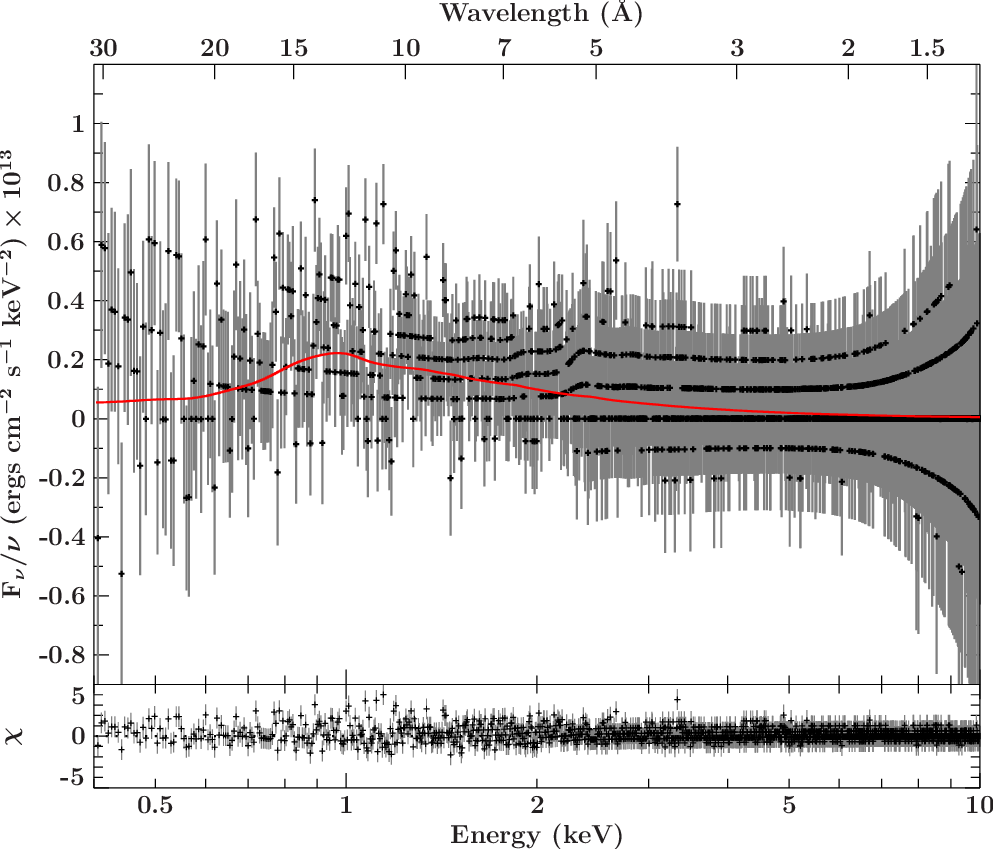}
\end{center}
\caption{The \textit{XMM-Newton} EPIC-pn observation of RT\,Cru fitted to the spectral model \textsf{tbnew}\,$\times$\,(\textsf{apec} $+$ \textsf{powerlaw}) plotted by a red color line. 
\label{rtcru:fig:xmm:model}
}
\end{figure}

To reproduce the \textit{XMM-Newton} PCA component, we assumed the spectral model \textsf{constant}\,$\times$\,\textsf{tbnew}\,$\times$\,(\textsf{apec} $+$ \textsf{powerlaw}). The energy-independent factor in the \textsc{xspec} component \textsf{constant} was adjusted to obtain total counts similar to the EPIC-pn observation for the same exposure (see Table~\ref{rtcru:obs:log}). To better create the absorption curvature in the soft X-ray excess, we employed the \textsf{tbnew} component \citep{Wilms2000}. Variability in the simulated 10\,ks-segmented spectra was made via multiplying the normalization factor of the \textsc{xspec} component \textsf{apec} by $1 + A_{1}(t)$, where $A_{1}(t)$ is the time series of the PCA component obtained from the \textit{XMM-Newton} observation (see Fig.\,\ref{rtcru:fig:pca:1}). 
The default values of the model parameters were determined from spectral analysis of the \textit{XMM-Newton} observation in \textsc{isis} using the aforementioned spectral model without the \textsf{constant} component. The best-fitting values of the model parameters are given in Table~\ref{rtcru:model:xmm}, along with the confidence limits (90\%) obtained using the \textsc{isis} function \textsf{conf\_loop}. The best-fit spectral model of the \textit{XMM-Newton} data is shown in Fig.\,\ref{rtcru:fig:xmm:model}.

\begin{table}
\begin{center}
\caption{Best-fitting model parameters for the \textit{XMM-Newton} observation.  
\label{rtcru:model:xmm}
}
\begin{tabular}{lcc}
\hline\hline
\noalign{\smallskip}
{\textsc{xspec}} & {Parameter} & {Value}   \\
\noalign{\smallskip}
\hline 
\noalign{\smallskip}
\textsf{tbnew} & $N_{\rm H}(10^{21}\mathrm{cm}^{-2})$ \dotfill & $ {  5.98 }_{ -1.93 }^{ +2.61 } $    \\
\noalign{\smallskip}
\textsf{apec}   & $kT$ (keV) \dotfill &  $ {  0.92 }_{ -0.68 }^{  +0.48 } $  \\  
                 & $K_{\rm apc}$ ($10^{-6}$)  \dotfill & $ {  6.70 }_{ -4.68 }^{  +0.62 } $   \\
\noalign{\smallskip}    
\textsf{powerlaw}   & $\Gamma$ (keV) \dotfill &  $ {  1.88 }_{ -0.50 }^{  +0.66 } $  \\  
                 & $K_{\rm pl}$ ($10^{-5}$)  \dotfill & $ {  2.56 }_{ -0.96 }^{ +1.77 } $     \\
\noalign{\smallskip}
\hline
\noalign{\smallskip}
\end{tabular}
\end{center}
\end{table}

\begin{figure*}
\begin{center}
\includegraphics[width=0.65\textwidth, trim = 0 0 0 0, clip, angle=0]{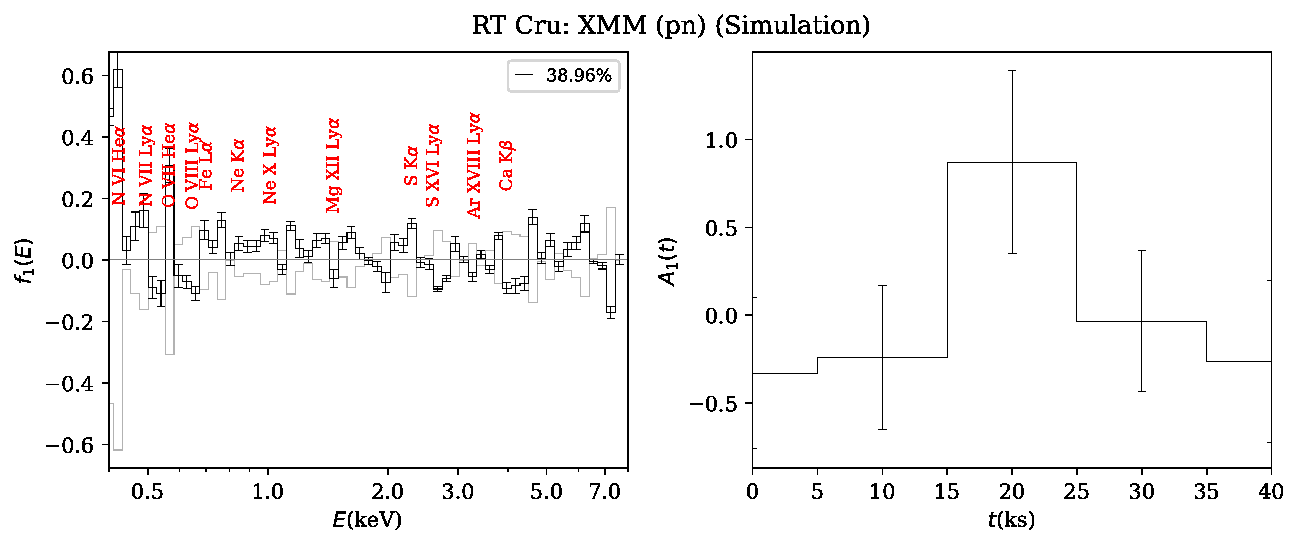}%
\includegraphics[width=0.3\textwidth, trim = 0 0 0 0, clip, angle=0]{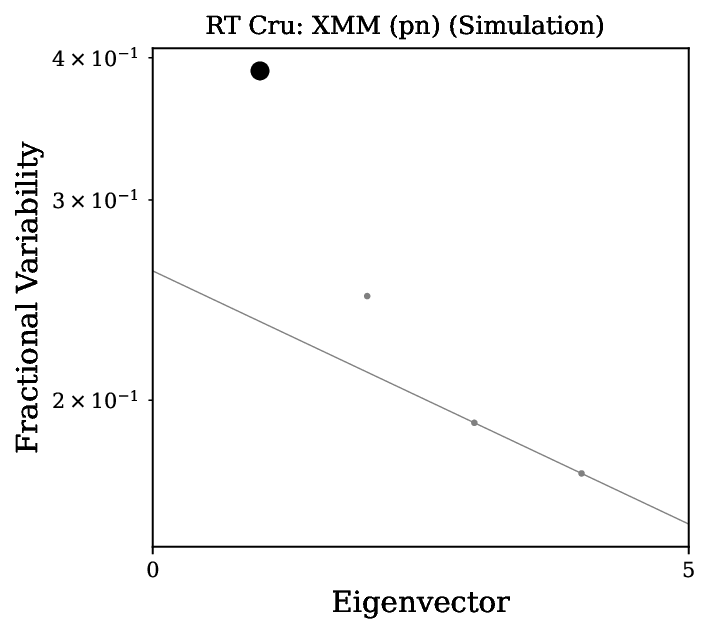}\\
\includegraphics[width=0.65\textwidth, trim = 0 0 0 0, clip, angle=0]{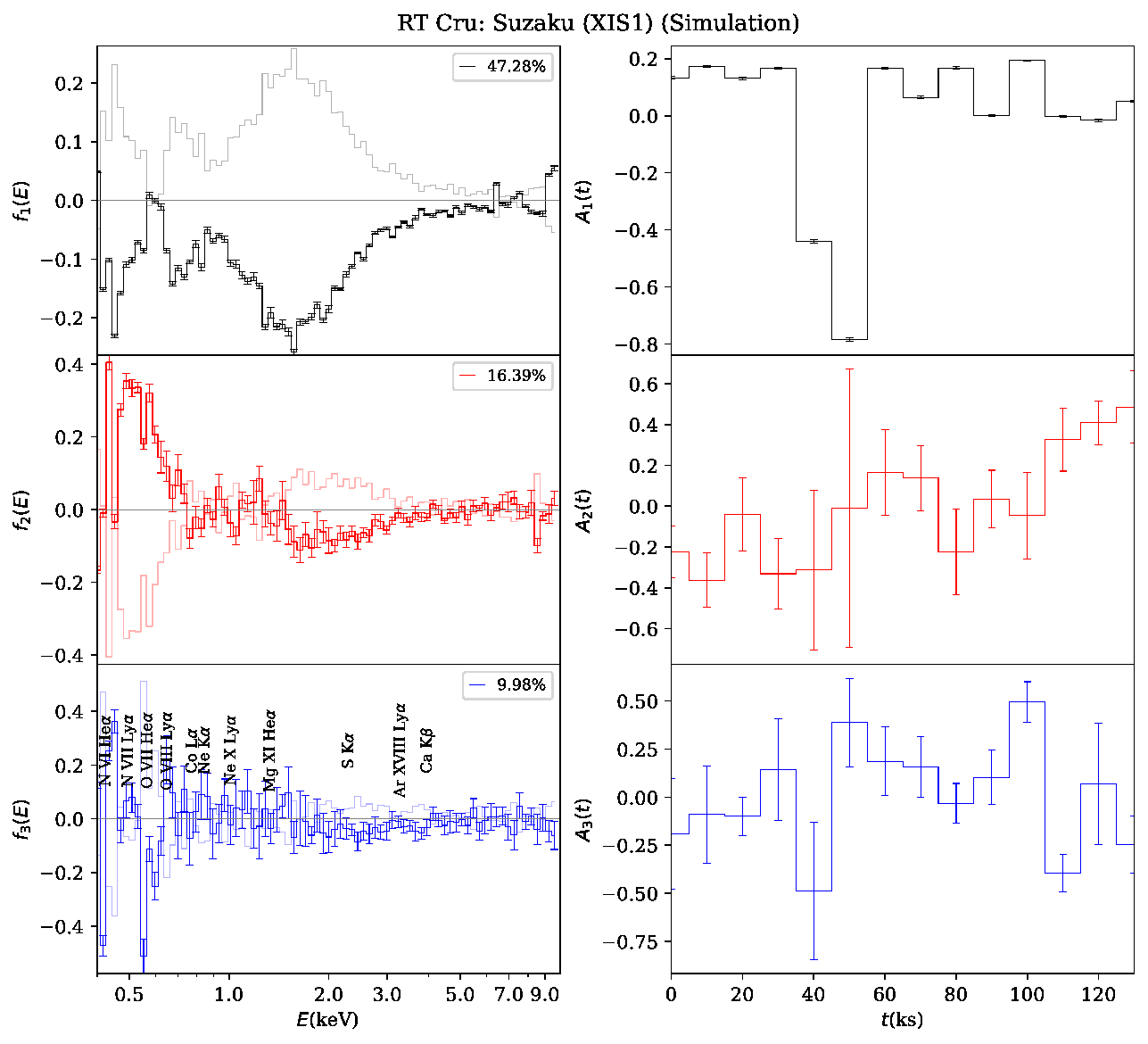}%
\includegraphics[width=0.3\textwidth, trim = 0 0 0 0, clip, angle=0]{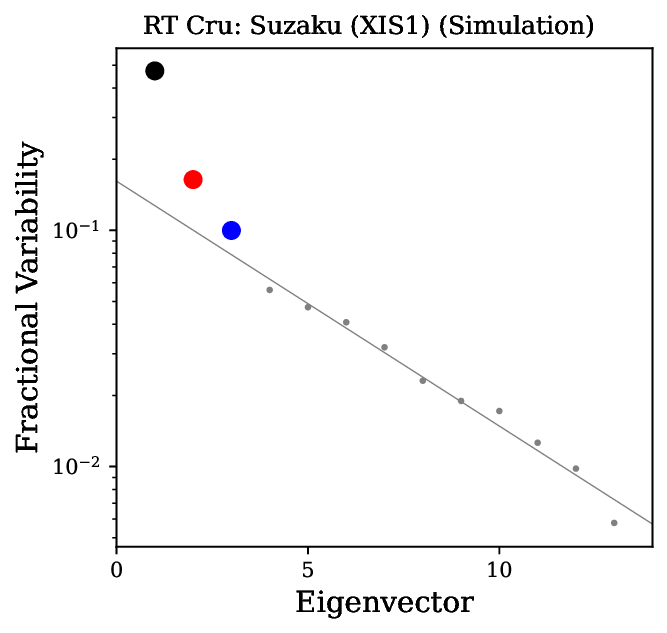}\\
\end{center}
\caption{The simulated PCA spectra $f_{i}(E)$ (left), the corresponding time series $A_{i}(t)$ (middle) and LEV diagram (right) of the spectra produced at intervals of 10\,ks with the \textit{XMM-Newton}/EPIC-pn and \textit{Suzaku}/XIS1 response data using the model parameters from Tables \ref{rtcru:model:xmm} and \ref{rtcru:model:param}, respectively, incorporating variations in the \textsf{apec} normalization factor into the \textit{XMM-Newton} and \textit{Suzaku} simulations, as well as alterations in the columns of the \textsf{pcfabs} components, and in the normalization factors of \textsf{compTT} and \textsf{apec} in the \textit{Suzaku} simulations, as described in the text.
\label{rtcru:fig:pca:sim}
}
\end{figure*}

To simulate the \textit{Suzaku}/XIS1 spectra, we adopted the phenomenological model, \textsf{constant}\,$\times$\,\textsf{pcfabs}\,$\times$\,(\textsf{diskbb} $+$ \textsf{apec}) $+$ \textsf{pcfabs}\,$\times$\,\textsf{compTT}, which is based on what was derived in \S\,\ref{rtcru:pca:construct}. The default values of the model parameters were set to those derived for PCA1 from spectral modeling listed in Table~\ref{rtcru:model:param}. Similarly, the \textsf{constant} component was used to reproduce the total counts akin to those seen in Table~\ref{rtcru:obs:log} for the same total exposures. Variations in the simulated 10\,ks-segmented spectra were then emulated by varying the column densities of the \textsf{pcfabs} components and the normalization factors of \textsf{compTT} and \textsf{apec}. Accordingly, we multiplied $N_{\rm H,dsk}$ and $N_{\rm H,cmp}$ of the \textsf{pcfabs} components by $1 + k_1 A_{1}(t)$, and set the normalization factors of the \textsc{xspec} functions \textsf{compTT} and \textsf{apec} to $[1 + k_2 A_{2}(t)] \times K_{\rm apc}$ and $[1 + k_3 A_{3}(t)] \times K_{\rm cmp}$, respectively, where $k_i$ are arbitrary constants for adjusting the variability fractions and $A_{i}(t)$ are the time series of the three PCA components derived from the \textit{Suzaku}/XIS1 observations plotted in Fig.\,\ref{rtcru:fig:pca:1}. The best-matched simulations were made with $k_1=0.9$, $k_2=2.4$, and $k_3=0.5$.

Figure~\ref{rtcru:fig:pca:sim} presents the PCA components deduced from the simulated \textit{XMM-Newton}/EPIC-pn and \textit{Suzaku}/XIS1 spectra. It can be seen that variable normalization of the \textsf{apec} component leads to the formation of a series of peaks in the PCA spectrum of the EPIC-pn simulation, which is comparable to $f_1(E)$ in Fig.\,\ref{rtcru:fig:pca:1} derived from the \textit{XMM-Newton} observation.
Moreover, the three statistically significant PCA components were obtained from the simulated XIS1 spectra made with variations in absorbing columns, Comptonization continua (\textsf{compTT}), and thermal emission (\textsf{apec}) scales. We see that the normalized PCA spectra $f_i(E)$ are roughly similar to those from the \textit{Suzaku}/XIS1 observations. 

Hence, our simulations suggest that the hardness spectral transition seen in the \textit{Suzaku}/XIS1 observations  could be mostly due to varying absorbing columns and partially caused by some variations in the continuum. Furthermore, our results indicate that some changes in the thermal plasma emission could contribute to the PCA component of the \textit{XMM-Newton} data and the third PCA component obtained from the \textit{Suzaku}/XIS1 observations. 

\section{Discussions}
\label{rtcru:discussions}

\subsection{Transient Nature}

The hard X-ray source IGR\,J12349-6434 was first detected with the IBIS instrument aboard INTEGRAL in 2003--2004 \citep{Chernyakova2005}. Its association with RT\,Cru was suggested by \citet{Masetti2005} and confirmed by Swift observations \citep{Tueller2005a}. Although the source was at a level of $\sim 3$\,mCrab in the 20--60\,keV energy band during the first detection \citep{Chernyakova2005}, it was seen 3.4 times brighter at 13\,mCrab in the 18--40\,keV energy band in 2012 \citep{Sguera2012}, followed by a flux level of $\sim 6$\,mCrab in 2015 \citep{Sguera2015}. The ASAS and AAVSO optical light curves also indicated that RT\,Cru became brighter from 13.5 to 11.3\,mag between 1998 and 2001; there was then a gradual decrease in the brightness to 12.1\,mag in 2006, and later an increase to 11.8\,mag in 2009 and 11.5\,mag in 2012, followed by a decrease to 12.6\,mag in 2017 \citep{Luna2018}. The \textit{Swift}/BAT survey revealed that the source reached its highest X-ray brightness between 2011 and 2012 coincident with the optical peak in 2012 \citep{Luna2018}. As seen in Fig.~\ref{rtcru:fig:lc}, the \textit{Suzaku} hardness ratios exhibited harder X-rays in 2012 than in 2007. According to the spectral fitting results by \citet[][]{Luna2018}, this transition could be related to a later increase in the column density of the absorbing material, as well as a rise in the X-ray brightness owing to higher accretion rates.

The most conspicuous finding of our PCA is the appearance of the absorbing column and the continuum in the first and second principal components of the \textit{Suzaku} XIS1 data. The spectral analysis of the \textit{Suzaku} data by \cite{Luna2018} provided an absorbing column of $1.4\times 10^{22}$ and $12 \times 10^{22}$\,cm$^{-2}$ for the XIS1 and PIN observations in 2007, much lower than $4.8\times 10^{22}$ and $28 \times 10^{22}$\,cm$^{-2}$ in 2012, respectively. In particular, as seen in Fig.\,\ref{rtcru:fig:pca:1}, this absorption has the strongest effect at energies between 1 and 4\,keV. 

\subsection{Flickering Behavior}

Binary systems hosting degenerate cores usually manifest flickering, which is associated with accretion physics \citep[see, e.g.,][]{Luna2013,Merc2024}. RT\,Cru is characterized by flickering, referring to the stochastic variability in the light curve characterized by less than tenths of a magnitude, occurring on scales from seconds to minutes. The flickering behavior in this object has been recorded from the optical band \citep{Cieslinski1994} to UV \citep{Luna2018} and X-rays \citep{Ducci2016,Danehkar2021}. This behavior was further seen in the $B$, $V$, and $R$ bands from the ground-based observations and photometry made with the NASA TESS mission \citep{Pujol2023}. Its TESS light curves show accumulation-induced flickering variability on timescales of minutes, with substantial variation in flickering \citep{Merc2024}. Our variability simulations suggest that changes in the column density or/and covering fraction of the absorbing material, along with the X-ray brightness from accretion processes, could potentially lead to long-term variations in the soft excess below 4\,keV, which aligns with the spectral fitting results of \citet{Luna2018}. Moreover, simulated X-ray spectra indicate that rapid flickering-type variations might be caused by some changes in the thermal plasma emission, which  \citet{Luna2007} found likely originates from a boundary layer of the accretion disk around a massive white dwarf rather than a magnetically channeled flow.

The disappearance of optical Balmer emission lines and decreases in $U$, $B$, and $V$ flickering amplitudes were also recorded in RT\,Cru in 2019, attributed to a decline in the accretion process, followed by its reappearance in the later years associated with restoring the accretion flow \citep{Pujol2023}. In particular, flickering in the $\delta$-type symbiotic star T\,CrB is predominantly detected in hard X-rays during two active phases with periods of $\sim 1000$d and $\sim 5000$d, and seems to be produced in the boundary layer due to variable mass transfer \citep{Ilkiewicz2016}. Moreover, flickering in the $\beta$-class symbiotic star AG Peg, whose soft emission may be a result of CSWs, resembles those produced by accretion processes in the X-ray spectra over 2013--2015 \citep{Zhekov2016}. 
Rapid, low-amplitude UV flickering has been seen in $\delta$ and $\beta$/$\delta$ sources, which seem to originate from accretion processes rather than quasi-steady thermonuclear burning on the white dwarf surface, or CSWs \citep{Luna2013}.

\section{Conclusion}
\label{rtcru:conclusion}

We have used hardness ratio and principal component analysis to assess the spectral variability of the $\delta$-type symbiotic binary RT\,Cru seen in the archival multi-mission data collected with various X-ray telescopes. Our key results are as follows:

(i) Our hardness ratio analysis revealed that both the soft (0.4--1.1\,keV) and hard (2.6--10\,keV) excesses of the source in 2012 are stronger than those in 2007, according to the \textit{Suzaku} observations over the two epochs. Moreover, hourly flickering variations are seen in the \textit{XMM-Newton} and \textit{NuSTAR} light curves in Fig.\,\ref{rtcru:fig:lc}, which contribute to the stochastic variability. As seen in Fig.\,\ref{rtcru:fig:hdr}, the long-term spectral transition between the two \textit{Suzaku} data sets is much more predominant than those made by hourly variations.

(ii) Our statistical analysis reveals that the source experienced statistically significant variations over the full band of the \textit{XMM-Newton} data, albeit with homogeneous hardness states, as indicated by the von Neumann ratios ($\eta$) and the normality tests. Some variations were also observed in both the \textit{Suzaku}/XIS1 observations, the first \textit{Suzaku}/PIN data, and the \textit{NuSTAR} data, as confirmed by the von Neumann statistics. Furthermore, the A--D and S--W normality tests confirm that the X-ray source underwent a prolonged transition in its spectral characteristics between the 2007 and 2012 observations with \textit{Suzaku}. This finding is supported by the results of normality tests conducted on HR$_{1}$, HR$_{2}$, and HR$_{3}$, in addition to the presence of some stochastic variations in HR$_{1}$ of the second \textit{Suzaku}/XIS1 dataset and HR$_{3}$ of the \textit{NuSTAR} observation.

(iii) The primary PCA components derived from the time-stacked \textit{Chandra} HRC-S/LETG and \textit{XMM-Newton} EPIC-pn data likely suggest the presence of some thermal emission lines from H-like and He-like ions, with strong features in the soft excess, such as \ionic{N}{vii} Ly$\alpha$, \ionic{N}{vi} He$\alpha$, \ionic{O}{viii} Ly$\alpha$, and \ionic{O}{vii} He$\alpha$ (see Fig.\,\ref{rtcru:fig:pca:1}). Our simulated \textit{XMM-Newton} spectra imply that changes in the amplitudes of the thermal emission lines could lead to the similar PCA spectrum. 

(iv) PCA of the two \textit{Suzaku} XIS1 observations provides us with the three spectral components. The first PCA spectrum $f_{1}(E)$ likely corresponds to the line-of-sight absorbing material since the corresponding light curve $A_{1}(t)$ rises with decreases in the brightness $S+M+H$ as seen in Fig.\,\ref{rtcru:fig:pca:1}. The second component $f_{2}(E)$ is probably associated with an absorbed continuum, consisting of a soft blackbody-like spectrum and a hard spectrum. Based on our simulations of \textit{Suzaku} spectra, the X-ray variations are mostly caused by changes in the absorbing columns, with some effects also owing to alterations in the continuum. The last PCA spectrum $f_{3}(E)$ contains some spectral features akin to those seen in $f_{1}(E)$ of the \textit{XMM-Newton} data, so it may be related to the heavily obscured, soft plasma emission with $\sim 1.3$\,keV suggested by \citet{Danehkar2021} using low-count Bayesian statistics. Additionally, a similar collisional plasma temperature is derived from a phenomenological model fitted to the reconstructed spectra of the first and second PCA components of the \textit{Suzaku} XIS1 data, as well as a simple model matched to the \textit{XMM-Newton} observation. Such a thermal emission feature can be created with a wind velocity of 1000 km\,s$^{-1}$ similar to jets found in CH\,Cyg \citep{Karovska2007,Karovska2010}. Our simulations of the XIS1 spectra also illustrate that some changes in the thermal plasma emission could lead to the third PCA component of \textit{Suzaku}, resulting in flickering-type variations.

(v) The \textit{Suzaku} HXD-PIN and \textit{NuSTAR} observations show a powerlaw-like continuum, but there is no separate component for the absorbing column in the hard excesses. This may imply that the absorbing material mainly obscures the spectrum below 4\,keV, as evidenced by $f_{1}(E)$ of the the \textit{Suzaku} XIS1 data in Fig.\,\ref{rtcru:fig:pca:1}. 
Moreover, PCA of the \textit{NuSTAR} data also offers a second component that may contain some emission features similar to those seen in $f_{1}(E)$ of the \textit{XMM-Newton} data and $f_{3}(E)$ of the \textit{Suzaku} XIS, with the energies corresponding to \ionic{Ar}{xviii} Ly$\alpha$ 3.3\,keV, Ca K$\alpha$ 3.7\,keV, and Ca K$\beta$ 4\,keV. However, we should caution that \textit{NuSTAR} has a spectral resolution much lower than those of the \textit{XMM-Newon} and \textit{Suzaku} XIS, which reduces the reliability of those line features seen in the \textit{NuSTAR} PCA spectrum. 

(vi) Finally, our PCA study implies that the primary factor contributing to the significant hardness transition observed across the two \textit{Suzaku} data sets (Fig.\,\ref{rtcru:fig:hdr}) is mainly made by changes in the absorbing material and partially because of the source continuum. This might be associated with the fact that hard X-ray flux significantly decreased in 2019, which could have been because less material was accreting into the degenerate white dwarf as proposed by \citet{Pujol2023}. Moreover, we found no resolved PCA component for the obscuring material in other telescopes since it was not largely variable over the course of those observations. Additionally, it seems that the absorbing column may primarily affect the soft excess ($<3$keV), which is not covered by the energy ranges of the \textit{Suzaku} PIN and \textit{NuSTAR}.

In summary, our eigenvector-based multivariate analysis of the \textit{Suzaku} data of RT\,Cru suggests that the hardness transition seen over two epochs is related primarily to changes in the absorbing material and partially to the X-ray brightness caused by accretion processes. In addition, our analysis of the multi-mission X-ray data taken with different telescopes supports the likelihood of a soft thermal plasma emission component, as previously proposed by \citet{Danehkar2021} with Bayesian analysis, which may be heavily obscured by the line-of-sight absorbing material as well as high levels of background noise. A recent statistical approach by \citet{Zhang2023} could not robustly constrain any soft-band emission lines, but it did provide an upper confidence level of 1\,keV for the plasma temperature. Furthermore, changes in thermal plasma emission from an accretion disk boundary layer likely cause the flickering-type variations observed in this object. Some future X-ray telescopes, such as the proposed \textit{Arcus} \citep{Smith2016}, with a much higher sensitivity below 1\,keV, will be able to capture the emission features in the soft band. Moreover, future high-spectral resolution observations with the recently launched telescope \textit{XRISM} \citep{Tashiro2020}  will certainly help us disclose further details of the X-ray spectral features of RT\,Cru.



\begin{acknowledgments}
We thank the anonymous referee for useful comments. A.D. thanks the ADAP grant 80NSSC22K0626 from NASA Headquarters (HQ) and the award 80NSSC23K1098 from Goddard Space Flight Center (GSFC). G.J.M.L. is a member of the CIC-CONICET (Argentina) and acknowledges support from grant ANPCYT-PICT 0901/2017. Based on observations obtained with the \textit{Suzaku} satellite, a joint mission of the JAXA and NASA space agencies; \textit{XMM-Newton}, an ESA science mission with instruments and contributions directly funded by ESA Member States and NASA; the \textit{NuSTAR} mission, a project led by Caltech, managed by JPL, funded by NASA; and
the \textit{Chandra} X-ray Observatory (CXO). This research made use of of data and/or software provided by the High Energy Astrophysics Science Archive Research Center \citep{NHEASARC2014}; software provided by the \textit{Chandra} X-ray Center in the application packages \textsc{ciao} and \textsf{Sherpa}; the \textsc{sas} software for XMM-Newton; a collection of \textsc{isis} functions (ISISscripts) provided by ECAP/Remeis Observatory and MIT; and \textsf{Astropy}, a community-developed core Python package for Astronomy.

\end{acknowledgments}



%

\vspace{1mm}


\facilities{CXO (HRC, LETG), \textit{XMM} (pn), \textit{NuSTAR} (FPMA, FPMB), \textit{Suzaku} (XIS, PIN).}


\software{\textsf{NumPy} \citep{Harris2020}, \textsf{SciPy} \citep{Virtanen2020}, \textsf{Matplotlib} \citep{Hunter2007}, \textsf{Astropy} \citep{AstropyCollaboration2013}.}


{ \small 
\begin{center}
\textbf{ORCID iDs}
\end{center}
\vspace{-5pt}

\noindent A.~Danehkar \orcidauthor{0000-0003-4552-5997} \url{https://orcid.org/0000-0003-4552-5997}

\noindent J.~J.~Drake \orcidauthor{0000-0002-0210-2276} \url{https://orcid.org/0000-0002-0210-2276}

\noindent G.~J.~M.~Luna \orcidauthor{0000-0002-2647-4373} \url{https://orcid.org/0000-0002-2647-4373}
}





\end{document}